\documentclass[11pt]{article}

\pdfoutput=1
\usepackage{ifpdf}
\ifpdf

\newcommand{\gea}{\raisebox{-.3ex}{\small $ \
\stackrel{\textstyle >}{\sim} $ }}
\newcommand{\bbox}[1]{\mbox{\boldmath $#1$}}
\usepackage{graphics}
\def\be{\begin{eqnarray}}
\def\ee{\end{eqnarray}}
\begin{document}

\title{\bf Nuclear Forces from Chiral Effective Field
Theory\thanks{Lecture series presented at the DAE-BRNS Workshop
on Physics and Astrophysics of Hadrons and Hadronic
Matter, Visva Bharati University, Santiniketan, West Bengal,
India, November 2006.}}
\author{R.Machleidt\\
Department of Physics, University of Idaho, Moscow, Idaho, U.S.A.}
\date{\today}
\maketitle

\begin{abstract}
In this lecture series, I present 
the recent progress in our understanding of nuclear forces
in terms of chiral effective field theory.
\end{abstract}

\tableofcontents

\newpage

\section{Introduction and Historical Perspective}

The theory of nuclear forces has a long history 
(cf.~Table~\ref{tab_hist}).
Based upon the seminal idea by Yukawa~\cite{Yuk35}, 
first field-theoretic attempts
to derive the nucleon-nucleon ($NN$) interaction
focused on pion-exchange.
While the one-pion exchange turned out to be very useful
in explaining $NN$ scattering data and the properties
of the deuteron~\cite{Sup56}, 
multi-pion exchange was beset with
serious ambiguities ~\cite{TMO52,BW53}.
Thus, the ``pion theories'' of the 1950s
are generally judged as failures---for reasons
we understand today: pion dynamics is constrained by chiral
symmetry, a crucial point that was unknown in the 1950s.

Historically, the experimental discovery of heavy 
mesons~\cite{Erw61} in the early 1960s
saved the situation. The one-boson-exchange (OBE)
model~\cite{OBEP,Mac89} emerged which is still the most economical
and quantitative
phenomenology for describing the 
nuclear force~\cite{Sto94,Mac01}.
The weak point of this model, however, is the scalar-isoscalar
``sigma'' or ``epsilon'' boson, for which the empirical
evidence remains controversial. Since this boson is associated
with the  correlated (or resonant) exchange of two pions,
a vast theoretical effort that occupied more than a decade 
was launched to derive the 2$\pi$-exchange contribution
to the nuclear force, which creates the intermediate 
range attraction.
For this, dispersion theory as well as 
field theory were invoked producing the
Stony Brook~\cite{JRV75},
Paris~\cite{Vin79,Lac80}, and
Bonn~\cite{Mac89,MHE87}
potentials.

The nuclear force problem appeared to be solved; however,
with the discovery of quantum chromodynamics (QCD), 
all ``meson theories'' were
relegated to models and the attempts to derive
the nuclear force started all over again.

The problem with a derivation from QCD is that
this theory is non-perturbative in the low-energy regime
characteristic of nuclear physics, which makes direct solutions
impossible.
Therefore, during the first round of new attempts,
QCD-inspired quark models~\cite{MW88} became popular. 
These models are able to reproduce
qualitatively and, in some cases, semi-quantitatively
the gross features of the nuclear 
force~\cite{EFV00,Wu00}.
However, on a critical note, it has been pointed out
that these quark-based
approaches are nothing but
another set of models and, thus, do not represent any
fundamental progress. Equally well, one may then stay
with the simpler and much more quantitative meson models.

\begin{table}[t]
\caption{Seven Decades of Struggle:
The Theory of Nuclear Forces
\label{tab_hist}} 
\smallskip
\begin{tabular*}{\textwidth}{@{\extracolsep{\fill}}cccc}
\hline
\hline
\noalign{\smallskip}
\\
   & \bf 1935   &
\bf Yukawa: Meson Theory 
\\
\\
\hline
\noalign{\smallskip}
     &      &
{\it The ``Pion Theories''}
\\
 & \bf 1950's &
One-Pion Exchange: o.k.
\\
  &         &
Multi-Pion Exchange: disaster
\\
\hline
\noalign{\smallskip}
  &         & 
Many pions $\equiv$ multi-pion resonances:
\\
 & \bf 1960's & 
{\boldmath $\sigma$, $\rho$, $\omega$, ...}
\\
  &         & 
The One-Boson-Exchange Model: success
\\
\hline
\noalign{\smallskip}
  &         &
Refined meson models, including
\\
 & \bf 1970's & 
sophisticated {\boldmath $2\pi$} exchange contributions
\\
  &         & 
(Stony Brook, Paris, Bonn)
\\
\hline
\noalign{\smallskip}
  &         & 
Nuclear physicists discover
\\
  & \bf 1980's &  
{\bf QCD}
\\
  &        & 
Quark Cluster Models
\\
\hline
\noalign{\smallskip}
  &        &
Nuclear physicists discover {\bf EFT}
\\
 & \bf 1990's &
Weinberg, van Kolck
\\
 & \bf and beyond &
{\bf Back to Pion Theory!}
\\
  &           &
{\it But, constrained by Chiral Symmetry: success}
\\
\hline
\hline
\end{tabular*}
\end{table}

A major breakthrough occurred when 
the concept of an effective field theory (EFT) was introduced
and applied to low-energy QCD.
As outlined by Weinberg in a seminal paper~\cite{Wei79},
one has to write down the most general Lagrangian consistent
with the assumed symmetry principles, particularly
the (broken) chiral symmetry of QCD.
At low energy, the effective degrees of freedom are pions and
nucleons rather than quarks and gluons; heavy mesons and
nucleon resonances are ``integrated out''.
So, the circle of history is closing and we are back 
to Yukawa's meson theory,
except that we have learned to add one important refinement to the theory:
broken chiral symmetry is a crucial constraint that generates
and controls the dynamics and establishes a clear connection
with the underlying theory, QCD.

Following the first initiative by Weinberg \cite{Wei90}, pioneering
work was performed by Ord\'o\~nez, Ray, and
van Kolck \cite{ORK94,Kol99} who 
constructed a $NN$ potential in coordinate space
based upon chiral perturbation theory at
next-to-next-to-leading order.
The results were encouraging and
many researchers
became attracted to the new
field~\cite{CPS92,KBW97,KGW98,EGM98,EM02a,EM02,EM03}. 
As a consequence, nuclear EFT has developed into one of the most popular
branches of modern nuclear physics~\cite{ME05,BK02}.

It is the purpose of these lectures to describe
in some detail
the recent progress in our understanding of nuclear
forces in terms of nuclear EFT.

\section{QCD and the Nuclear Force}

Quantum chromodynamics (QCD) is the theory of strong interactions.
It deals with quarks, gluons and their interactions and is
part of the Standard Model of Particle Physics.
QCD is a non-Abelian gauge field theory
with color $SU(3)$ the underlying gauge group.
The non-Abelian nature of the theory has dramatic
consequences. While 
the interaction between colored objects is weak 
at short distances or high momentum transfer
(``asymptotic freedom'');
it is strong at long distances ($\gea 1$ fm) or low energies,
leading to the confinement of quarks into colorless
objects, the hadrons. Consequently, QCD allows for a 
perturbative analysis at large energies, whereas it is
highly non-perturbative in the low-energy regime.
Nuclear physics resides at low energies and
the force between nucleons is
a residual QCD interaction. 
Therefore, in terms of quarks and gluons, the nuclear force
is a very complicated problem.

\section{Effective Field Theory for Low-Energy QCD
\label{sec_EFT}}

The way out of the dilemma 
of how to derive the nuclear force from QCD
is provided by the 
effective field theory (EFT) concept. First, one needs to identify
the relevant degrees of freedom. For the ground state and the
low-energy excitation spectrum of
an atomic nucleus as well as for conventional nuclear
reactions,
quarks and gluons are ineffective degrees of freedom,
while nucleons and pions are the appropriate ones.
Second; to make sure that this EFT is not just another phenomenology,
the EFT must observe
all relevant symmetries of the underlying theory.
This requirement is based upon a `folk theorem' by
Weinberg~\cite{Wei79}:
\begin{quote}
If one writes down the most general possible Lagrangian, including {\it all}
terms consistent with assumed symmetry principles,
and then calculates matrix elements with this Lagrangian to any given order of
perturbation theory, the result will simply be the most general possible 
S-matrix consistent with analyticity, perturbative unitarity,
cluster decomposition, and the assumed symmetry principles.
\end{quote}
Thus, the EFT program consists of the following steps:
\begin{enumerate}
\item
Identify the degrees of freedom relevant at the resolution
scale of (low-energy) nuclear physics: nucleons and pions.
\item
Identify the relevant symmetries of low-energy QCD and
investigate if and how they are broken.
\item
Construct the most general Lagrangian consistent with those
symmetries and the symmetry breaking.
\item
Design an organizational scheme that can distinguish
between more and less important contributions: 
a low-momentum expansion.
\item
Guided by the expansion, calculate Feynman diagrams
to the the desired accuracy for the problem under consideration.
\end{enumerate}
We will now elaborate on these steps, one by one.

\subsection{Symmetries of Low-Energy QCD}

In this section, we will give a brief introduction into 
(low-energy) QCD,
its symmetries and symmetry breaking.
A more detailed introduction can be found
in the excellent lecture series by Scherer 
and Schindler~\cite{SS05}.

\subsubsection{Chiral Symmetry}

The QCD Lagrangian reads
\begin{equation}
{\cal L}_{\rm QCD} = \bar{q} (i \gamma^\mu {\cal D}_\mu
 - {\cal M})q
 - \frac14 
{\cal G}_{\mu\nu,a}
{\cal G}^{\mu\nu}_{a} 
\label{eq_LQCD}
\end{equation}
with the gauge-covariant derivative
\begin{equation}
{\cal D}_\mu = \partial_\mu +ig\frac{\lambda_a}{2}
{\cal A}_{\mu,a}
\end{equation}
and the gluon field strength tensor
\begin{equation}
{\cal G}_{\mu\nu,a} =
\partial_\mu {\cal A}_{\nu,a}
-\partial_\nu {\cal A}_{\mu,a}
 - g f_{abc}
{\cal A}_{\mu,b}
{\cal A}_{\nu,c} \,.
\end{equation}
In the above, $q$ denotes the quark fields and ${\cal M}$
the quark mass matrix. Further, $g$ is the
strong coupling constant and ${\cal A}_{\mu,a}$
are the gluon fields. The $\lambda_a$ are the
Gell-Mann matrices and the $f_{abc}$
the structure constants of the $SU(3)_{\rm color}$
Lie algebra $(a,b,c=1,\dots ,8)$;
summation over repeated indices is always implied.
The gluon-gluon term in the last equation arises
from the non-Abelian nature of the gauge theory
and is the reason for the peculiar features
of the color force.

On a typical hadronic scale, i.e., on a scale of low-mass
hadrons which are not Goldstone bosons, e.g., 
$m_\rho=0.78 \mbox{ GeV} \approx 1 \mbox{ GeV}$; 
the masses of the up $(u)$, down $(d)$, and---to a certain
extend---strange (s) quarks are small~\cite{PDG}:
\begin{eqnarray}
m_u &=& 2\pm 1 \mbox{ MeV} 
\label{eq_umass} \\
m_d &=& 5\pm 2 \mbox{ MeV} 
\label{eq_dmass} \\
m_s &=& 95\pm 25 \mbox{ MeV}
\label{eq_smass}
\end{eqnarray}
It is therefore of interest to discuss the
QCD Lagrangian in the limit of vanishing quark
masses:
\begin{equation}
{\cal L}_{\rm QCD}^0 = \bar{q} i \gamma^\mu {\cal D}_\mu
q - \frac14 
{\cal G}_{\mu\nu,a}
{\cal G}^{\mu\nu}_{a} \,.
\end{equation}
Defining right- and left-handed quark fields,
\begin{equation}
q_R=P_Rq \,, \;\;\;
q_L=P_Lq \,,
\end{equation}
with 
\begin{equation}
P_R=\frac12(1+\gamma_5) \,, \;\;\;
P_L=\frac12(1-\gamma_5) \,,
\end{equation}
we can rewrite the Lagrangian as follows:
\begin{equation}
{\cal L}_{\rm QCD}^0 = 
\bar{q}_R i \gamma^\mu {\cal D}_\mu q_R 
+\bar{q}_L i \gamma^\mu {\cal D}_\mu q_L 
- \frac14 
{\cal G}_{\mu\nu,a}
{\cal G}^{\mu\nu}_{a} \, .
\end{equation}
Restricting ourselves now to
up and down quarks,
we see that
${\cal L}_{\rm QCD}^0$ 
is invariant under the global unitary transformations
\begin{equation}
q_R =
\left( \begin{array}{c}
u_R \\ d_R
\end{array} \right)
\longmapsto
\exp
\left(
-i \Theta_i^R \frac{\tau_i}{2}
\right)
\left( \begin{array}{c}
u_R \\ d_R
\end{array} \right) 
\end{equation}
and
\begin{equation}
q_L =
\left( \begin{array}{c}
u_L \\ d_L
\end{array} \right)
\longmapsto
\exp
\left(
-i \Theta_i^L \frac{\tau_i}{2}
\right)
\left( \begin{array}{c}
u_L \\ d_L
\end{array} \right) 
\,,
\end{equation}
where $\tau_i \; (i=1,2,3)$ 
are the generators of $SU(2)_{\rm flavor}$,
the usual Pauli spin matrices. 
{\it The right- and left-handed components of
massless quarks do not mix.} 
This is
$SU(2)_R\times SU(2)_R$ 
symmetry, also known as {\it chiral symmetry}.
Noether's Theorem implies the existence of
six conserved currents;
three right-handed currents
\begin{equation}
R^\mu_i = \bar{q}_R \gamma^\mu \frac{\tau_i}{2} q_R 
\;\;\;\; \mbox{\small\rm with} \;\;\;\; \partial_\mu R^\mu_i = 0
\end{equation}
and three left-handed currents
\begin{equation}
L^\mu_i = \bar{q}_L \gamma^\mu \frac{\tau_i}{2} q_L 
\;\;\;\; \mbox{\small\rm with} \;\;\;\; \partial_\mu L^\mu_i = 0
\,.
\end{equation}
It is useful to consider the following linear
combinations; namely,
three vector currents
\begin{equation}
V^\mu_i = R^\mu_i + L^\mu_i = \bar{q} \gamma^\mu \frac{\tau_i}{2} q
\;\;\;\; \mbox{\small\rm with} \;\;\;\; \partial_\mu V^\mu_i = 0
\end{equation}
and three axial-vector currents
\begin{equation}
A^\mu_i = R^\mu_i - L^\mu_i = \bar{q} \gamma^\mu \gamma_5 
\frac{\tau_i}{2} q
\;\;\;\; \mbox{\small\rm with} \;\;\;\; \partial_\mu A^\mu_i = 0
\,,
\end{equation}
which got their names from the fact that they
transform as vectors and axial-vectors, respectively.
Thus, the chiral $SU(2)_L \times SU(2)_R$ symmetry is equivalent to
$SU(2)_V \times SU(2)_A$, 
where the vector and axial-vector
transformations are given respectively by
\begin{equation}
q =
\left( \begin{array}{c}
u \\ d
\end{array} \right)
\longmapsto
\exp
\left(
-i \Theta_i^V \frac{\tau_i}{2}
\right)
\left( \begin{array}{c}
u \\ d
\end{array} \right) 
\end{equation}
and
\begin{equation}
q =
\left( \begin{array}{c}
u \\ d
\end{array} \right)
\longmapsto
\exp
\left(
-i \Theta_i^A \gamma_5 \frac{\tau_i}{2}
\right)
\left( \begin{array}{c}
u \\ d
\end{array} \right) \,.
\end{equation}
Obviously, the vector transformations are isospin rotations
and, therefore, invariance under vector transformations can be
identified with isospin symmetry.

There are the six conserved charges,
\begin{equation}
Q^V_i = \int d^3x \; V^0_i = \int d^3x \; q^\dagger (t,\vec x) 
\frac{\tau_i}{2} q(t,\vec x)
\;\;\;\; \mbox{\small\rm with} \;\;\;\; \frac{dQ^V_i}{dt} = 0
\end{equation}
and
\begin{equation}
Q^A_i = \int d^3x \; A^0_i = \int d^3x \; q^\dagger (t,\vec x) 
\gamma_5 \frac{\tau_i}{2} q(t,\vec x)
\;\;\;\; \mbox{\small\rm with} \;\;\;\; \frac{dQ^A_i}{dt} = 0
\, ,
\label{eq_charges}
\end{equation}
which are also generators of
$SU(2)_V \times SU(2)_A$.

\subsubsection{Explicit Symmetry Breaking}

The mass term  
 $- \bar{q}{\cal M}q$
in the QCD Lagrangian Eq.~(\ref{eq_LQCD}) 
breaks chiral symmetry explicitly. To better see this,
let's rewrite ${\cal M}$,
\begin{eqnarray}
{\cal M} & = & 
\left( \begin{array}{cc}
            m_u & 0 \\
              0  & m_d 
           \end{array} \right) \\
  & = & \frac12 (m_u+m_d) 
\left( \begin{array}{cc}
            1 & 0 \\
              0  & 1 
           \end{array} \right) 
+ \frac12 (m_u-m_d) 
\left( \begin{array}{cc}
            1 & 0 \\
              0  & -1 
           \end{array} \right) \\
 & = & \frac12 (m_u+m_d) \; I + \frac12 (m_u-m_d) \; \tau_3 \,.
\label{eq_mmatr}
\end{eqnarray}
The first term in the last equation in invariant under $SU(2)_V$
(isospin symmetry) and the second term vanishes for
$m_u=m_d$.
Thus, isospin is an exact symmetry if 
$m_u=m_d$.
However, both terms in Eq.~(\ref{eq_mmatr}) break $SU(2)_A$.
Since the up and down quark masses are small as compared to
the typical hadronic mass scale of $\approx 1$ GeV 
[cf.~Eqs.~(\ref{eq_umass}) and (\ref{eq_dmass})],
the explicit chiral symmetry breaking due to non-vanishing
quark masses is very small.

\subsubsection{Spontaneous Symmetry Breaking}

A (continuous) symmetry is said to be {\it spontaneously
broken} if a symmetry of the Lagrangian 
is not realized in the ground state of the system.
There is evidence that the chiral
symmetry of the QCD Lagrangian is spontaneously 
broken---for dynamical reasons of nonperturbative origin
which are not fully understood at this time.
The most plausible evidence comes from the hadron spectrum.
From chiral symmetry, one would naively expect the existence of 
degenerate hadron
multiplets of opposite parity, i.e., for any hadron of positive
parity one would expect a degenerate hadron state of negative 
parity and vice versa. However, these ``parity doublets'' are
not observed in nature. For example, take the $\rho$-meson,
a vector meson with negative parity ($1^-$) and mass 
776 MeV. There does exist a $1^+$ meson, the $a_1$, but it
has a mass of 1230 MeV and, thus, cannot be perceived
as degenerate with the $\rho$. On the other hand, the $\rho$
meson comes in three charge states (equivalent to
three isospin states), the $\rho^\pm$ and the $\rho^0$
with masses that differ by at most a few MeV. In summary,
in the QCD ground state (the hadron spectrum)
$SU(2)_V$ (isospin symmetry) is well observed,
while $SU(2)_A$ (axial symmetry) is broken.
Or, in other words,
$SU(2)_V\times SU(2)_A$ is broken down to $SU(2)_V$.

A spontaneously broken global symmetry implies the existence
of (massless) Goldstone bosons with the quantum numbers
of the broken generators. The broken generators are the $Q^A_i$
of Eq.~(\ref{eq_charges}) which are pseudoscalar.
The Goldstone bosons are identified with the isospin
triplet of the (pseudoscalar) pions, 
which explains why pions are so light.
The pion masses are not exactly zero because the up
and down quark masses
are not exactly zero either (explicit symmetry breaking).
Thus, pions are a truly remarkable species:
they reflect spontaneous as well as explicit symmetry
breaking.

\subsection{Chiral Effective Lagrangians Involving Pions
\label{sec_Lpi} }

The next step in our EFT program is to build the most general
Lagrangian consistent with the (broken) symmetries discussed
above.
An elegant formalism for the construction of such Lagrangians
was developed by 
Callan, Coleman, Wess, and Zumino (CCWZ)~\cite{CCWZ}
who worked out
the group-theoretical foundations 
of non-linear realizations of chiral symmetry. 
The Lagrangians given below are built upon the CCWZ
formalism.

As discussed, the relevant degrees of freedom are
pions (Goldstone bosons) and nucleons.
Since the interactions of Goldstone bosons must
vanish at zero momentum transfer and in the chiral
limit ($m\rightarrow 0$), the low-energy expansion
of the Lagrangian is arranged in powers of derivatives
and pion masses.
This is chiral perturbation theory (ChPT).

The Lagrangian consists of one part that deals with the
interaction among pions, ${\cal L}_{\pi\pi}$,
and another one that describes the interaction
between pions and the nucleon,
${\cal L}_{\pi N}$:
\begin{equation}
{\cal L}_{\rm eff} 
=
{\cal L}_{\pi\pi} 
+
{\cal L}_{\pi N} 
\end{equation}
with
\begin{equation}
{\cal L}_{\pi\pi} 
 = 
{\cal L}_{\pi\pi}^{(2)} 
 + {\cal L}_{\pi\pi}^{(4)} 
 + \ldots 
\end{equation}
and
\begin{equation}
{\cal L}_{\pi N} 
= 
{\cal L}_{\pi N}^{(1)} 
+
{\cal L}_{\pi N}^{(2)} 
+
{\cal L}_{\pi N}^{(3)} 
+ \ldots ,
\end{equation}
where the superscript refers to the number of derivatives or 
pion mass insertions (chiral dimension)
and the ellipsis stands for terms of higher dimension.

The {\it leading order (LO)} 
$\pi\pi$ Lagrangian is given by~\cite{GL84}
\begin{equation}
{\cal L}_{\pi\pi}^{(2)} =
\frac{f^2_\pi}{4} \, {\rm tr} \left[ \,
\partial^\mu U \partial_\mu U^\dagger 
+ m_\pi^2 ( U + U^\dagger)\, \right]
\label{eq_Lpipi}
\end{equation}
 and the LO relativistic $\pi N$ Lagrangian reads~\cite{GSS88}
\begin{equation}
{\cal L}^{(1)}_{\pi N}  =  
 \bar{\Psi} \left(i\gamma^\mu {D}_\mu 
 - M_N
 + \frac{g_A}{2} \gamma^\mu \gamma_5 u_\mu
  \right) \Psi  
\label{eq_piNrel}
\end{equation}
with
\begin{eqnarray}
{D}_\mu & = & \partial_\mu + \Gamma_\mu 
\\
\Gamma_\mu & = & \frac12      (
                           \xi^\dagger \partial_\mu \xi
                        +  \xi \partial_\mu \xi^\dagger)
= 
\frac{i}{4f^2_\pi} \,
\mbox{\boldmath $\tau$} \cdot 
 ( \mbox{\boldmath $\pi$}
\times
 \partial_\mu \mbox{\boldmath $\pi$})
+ \ldots
\\
u_\mu & = & i (
                           \xi^\dagger \partial_\mu \xi
                        -  \xi \partial_\mu \xi^\dagger)  
= -
\frac{1}{f_\pi} \,
\mbox{\boldmath $\tau$} \cdot 
 \partial_\mu \mbox{\boldmath $\pi$}
+ \ldots
\\
U & = & \xi^2 =
 1 + 
\frac{i}{f_\pi}
\mbox{\boldmath $\tau$} \cdot \mbox{\boldmath $\pi$}
-\frac{1}{2f_\pi^2} 
\mbox{\boldmath $\pi$}^2
-\frac{i\alpha}{f_\pi^3}
(\mbox{\boldmath $\tau$} \cdot \mbox{\boldmath $\pi$})^3
+\frac{8\alpha-1}{8f_\pi^4} 
\mbox{\boldmath $\pi$}^4
+ \ldots
\label{eq_alpha}
\end{eqnarray}
In Eq.~(\ref{eq_piNrel})
the chirally covariant derivative $D_\mu$ is applied
which introduces the ``gauge term'' $\Gamma_\mu$
(also known as chiral connection), a vector current
that leads to a coupling of pions with the nucleon.
Besides this, the Lagrangian includes a
coupling term which involves the 
axial vector $u_\mu$.
The $SU(2)$ matrix $U=\xi^2$ collects the Goldstone
pion fields.

In the above equations, $M_N$ denotes the nucleon mass,
$g_A$ the axial-vector coupling constant,
and $f_\pi$ the pion decay constant.
Numerical values will be given later.

The coefficient $\alpha$ that appears in Eq.~(\ref{eq_alpha})
is arbitrary. Therefore, diagrams with chiral vertices that
involve three or four pions must always be grouped together
such that the $\alpha$-dependence drops out
(cf.\ Fig.~\ref{fig_diag3}, below).

We apply
the heavy baryon (HB) formulation of 
chiral perturbation theory~\cite{BKM95}
in which the relativistic $\pi N$ Lagrangian is subjected
to an expansion in terms of powers of $1/M_N$ (kind of a
nonrelativistic expansion), the lowest order of which is 
\begin{eqnarray}
\widehat{\cal L}^{(1)}_{\pi N} & 
= & 
\bar{N} \left(
 i {D}_0 
 - \frac{g_A}{2} \; 
\vec \sigma \cdot \vec u
\right) N  
\nonumber \\
 & = & 
\bar{N} \left[ i \partial_0 
- \frac{1}{4f_\pi^2} \;
\mbox{\boldmath $\tau$} \cdot 
 ( \mbox{\boldmath $\pi$}
\times
 \partial_0 \mbox{\boldmath $\pi$})
- \frac{g_A}{2f_\pi} \;
\mbox{\boldmath $\tau$} \cdot 
 ( \vec \sigma \cdot \vec \nabla )
\mbox{\boldmath $\pi$} \right] N + \ldots
\label{eq_L1}
\end{eqnarray}
In the relativistic formulation, the nucleon
is represented
by a four-component Dirac spinor field, $\Psi$, 
while in the HB version, the nucleon, $N$,
is a Pauli spinor; in addition,
all nucleon fields include
Pauli spinors describing the isospin of the nucleon.

{\it At dimension two},
the relativistic $\pi N$ Lagrangian reads
\begin{equation}
{\cal L}^{(2)}_{\pi N} 
= \sum_{i=1}^{4} c_i \bar{\Psi} O^{(2)}_i \Psi \, .
\label{eq_L2rel}
\end{equation}
The various operators $O^{(2)}_i$ are given in Ref.~\cite{Fet00}.
The fundamental rule by which this Lagrangian---as well as all 
the other
ones---are assembled is that they must contain {\it all\/} terms
consistent with chiral symmetry and Lorentz invariance 
(apart from other trivial
symmetries) at a given chiral dimension (here: order two).
The parameters $c_i$ are known as low-energy constants (LECs)
and are determined empirically from fits to $\pi N$ data.

The HB projected $\pi N$ Lagrangian at order two 
is most conveniently broken up into two pieces,
\begin{equation}
\widehat{\cal L}^{(2)}_{\pi N} \, = \,
\widehat{\cal L}^{(2)}_{\pi N, \, \rm fix} \, + \,
\widehat{\cal L}^{(2)}_{\pi N, \, \rm ct} \, ,
\label{eq_L2}
\end{equation}
with
\begin{equation}
\widehat{\cal L}^{(2)}_{\pi N, \, \rm fix}  =  
 \bar{N} \left[
\frac{1}{2M_N}\: \vec D \cdot \vec D
+ i\, \frac{g_A}{4M_N}\: \{\vec \sigma \cdot \vec D, u_0\}
 \right] N
\label{eq_L2fix}
\end{equation}
and
\begin{eqnarray}
\widehat{\cal L}^{(2)}_{\pi N, \, \rm ct}
& = & 
 \bar{N} \left[
 2\,
c_1
\, m_\pi^2\, (U+U^\dagger)
\, + \, \left( 
c_2
- \frac{g_A^2}{8M_N}\right) u_0^2
 \, + \,
c_3
\, u_\mu  u^\mu
\right.  \nonumber \\ && \left.
+ \, \frac{i}{2} \left( 
c_4
+ \frac{1}{4M_N} \right) 
  \vec \sigma \cdot ( \vec u \times \vec u)
 \right] N \, .
\label{eq_L2ct}
\end{eqnarray}
Note that 
$\widehat{\cal L}^{(2)}_{\pi N, \, \rm fix}$  
is created entirely from the HB expansion of the relativistic
${\cal L}^{(1)}_{\pi N}$ and thus has no free parameters (``fixed''),
while 
$\widehat{\cal L}^{(2)}_{\pi N, \, \rm ct}$
is dominated by the new $\pi N$ contact terms proportional to the
$c_i$ parameters, besides some small $1/M_N$ corrections.

{\it At dimension three},
the relativistic $\pi N$ Lagrangian can be formally written as
\begin{equation}
{\cal L}^{(3)}_{\pi N} = \sum_{i=1}^{23} d_i \bar{\Psi} O^{(3)}_i \Psi \, ,
\label{eq_L3rel}
\end{equation}
with the operators, $O^{(3)}_i$, listed in Refs.~\cite{Fet00,FMS98}; 
not all 23 terms are of interest here.
The new LECs that occur at this order are the $d_i$.
Similar to the order two case,
the HB projected Lagrangian at order three can be broken into two pieces,
\begin{equation}
\widehat{\cal L}^{(3)}_{\pi N} \, = \,
\widehat{\cal L}^{(3)}_{\pi N, \, \rm fix} \, + \,
\widehat{\cal L}^{(3)}_{\pi N, \, \rm ct} \, ,
\label{eq_L3}
\end{equation}
with
$\widehat{\cal L}^{(3)}_{\pi N, \, \rm fix}$
and
$\widehat{\cal L}^{(3)}_{\pi N, \, \rm ct}$
given in Refs.~\cite{Fet00,FMS98}.

\subsection{Nucleon Contact Lagrangians}

Nucleon contact interactions consist of four nucleon
fields (four nucleon legs) and no meson fields.
Such terms are needed to renormalize
loop integrals, to make results reasonably independent
of regulators, and to parametrize the 
unresolved short-distance contributions to the nuclear
force. For more about contact terms,
see Sec.~\ref{sec_ct}.

Because of parity, nucleon contact interactions come only
in even numbers of derivatives, thus,
\begin{equation}
\label{eq_LNN}
{\mathcal L}_{NN} =
{\mathcal L}^{(0)}_{NN} +
{\mathcal L}^{(2)}_{NN} +
{\mathcal L}^{(4)}_{NN} + \ldots
\end{equation}

The lowest order (or leading order) $NN$ Lagrangian has no derivatives 
and reads~\cite{Wei90}
\begin{equation}
\label{eq_LNN0}
{\mathcal L}^{(0)}_{NN} =
-\frac{1}{2} C_S \bar{N} N \bar{N} N 
-\frac{1}{2} C_T \bar{N} \vec \sigma N \bar{N} \vec \sigma N \, ,
\end{equation}
where $N$ is the heavy baryon nucleon field.
$C_S$ and $C_T$ are unknown constants which are determined
by a fit to the $NN$ data.
The second order $NN$ Lagrangian is given by~\cite{ORK94}
\begin{eqnarray}
\label{eq_LNN2}
{\mathcal L}^{(2)}_{NN} &=&
-C'_1 [(\bar{N} \vec \nabla N)^2+ (\overline{\vec \nabla N} N)^2]
-C'_2 (\bar{N} \vec \nabla N)\cdot (\overline{\vec \nabla N} N)
\nonumber \\ &&
-C'_3 \bar{N} N [\bar N \vec \nabla^2 N+\overline{\vec \nabla^2 N} N]
\nonumber \\ &&
-i C'_4 [
\bar N \vec \nabla N \cdot 
(\overline{\vec \nabla N} \times \vec \sigma N) +
\overline{(\vec \nabla N)} N \cdot 
(\bar N \vec \sigma \times \vec \nabla N) ]
\nonumber \\ &&
-i C'_5 \bar N N(\overline{\vec \nabla N} \cdot \vec \sigma \times \vec \nabla N)
-i C'_6 (\bar N \vec \sigma N)\cdot (\overline{\vec \nabla N} \times \vec \nabla N)
\nonumber \\ &&
-(C'_7 \delta_{ik} \delta_{jl}+C'_8 \delta_{il} \delta_{kj}
+C'_9 \delta_{ij} \delta_{kl})
\nonumber \\ && \times
[\bar N \sigma_k \partial_i N \bar N \sigma_l \partial_j N +
\overline{\partial_i N} \sigma_k N \overline{\partial_j N} \sigma_l N ]
\nonumber \\ &&
-(C'_{10} \delta_{ik} \delta_{jl}+C'_{11} \delta_{il} \delta_{kj}+C'_{12} \delta_{ij} \delta_{kl})
\bar N \sigma_k \partial_i N \overline{\partial_j N} \sigma_l N
\nonumber \\ &&
-(\frac{1}{2} C'_{13} (\delta_{ik} \delta_{jl}+\delta_{il} \delta_{kj}) 
\nonumber \\ &&
+C'_{14} \delta_{ij} \delta_{kl})
[\overline{\partial_i N} \sigma_k \partial_j N + \overline{\partial_j N} \sigma_k \partial_i N]
\bar N \sigma_l N \, .
\end{eqnarray}
Similar to $C_S$ and $C_T$, the $C'_i$ are unknown constants 
which are fixed in a fit to the $NN$ data.
Obviously, these contact Lagrangians blow up quite a bit
with increasing order, which why we do not give
${\mathcal L}^{(4)}_{NN}$
explicitly here.

\section{Nuclear Forces from EFT: Overview}

In the beginning of Sec.~\ref{sec_EFT},
we spelled out the steps we have to take to accomplish
our EFT program for the derivation of nuclear forces.
So far, we discussed steps one to three.
What is left are steps four (low-momentum expansion) and
five (Feynman diagrams).
In this section, we will say more about the expansion
we are using and
give an overview of the Feynman diagrams that arise
order by order.

\subsection{Chiral Perturbation Theory and Power Counting}
In ChPT, we
analyze contributions
in terms of powers of small momenta over the large scale:
$(Q/\Lambda_\chi)^\nu$,
where $Q$ stands for a momentum (nucleon three-momentum or
pion four-momentum) or a pion mass and $\Lambda_\chi \approx 1$ GeV
is the chiral symmetry breaking scale (hadronic scale).
Determining the power $\nu$ at which a given diagram 
contributes has become known as power counting.
For a non-iterative contribution involving $A$ nucleons,
the power $\nu$ is given by
\begin{equation}
\nu = -2 +2A - 2C + 2L + \sum_i \Delta_i 
 \, , 
\end{equation}
with
\begin{equation}
\Delta_i  \equiv   d_i + \frac{n_i}{2} - 2  \, ,
\end{equation}
where $C$ denotes the number of separately connected pieces and
$L$ the number of loops in the diagram;
$d_i$ is the number of derivatives or pion-mass insertions 
and
$n_i$ the number of nucleon fields 
involved in vertex $i$;
the sum runs over all vertices contained in the diagram 
under consideration.
Note that for an irreducible $NN$ diagram ($A=2$), the above
formula reduces to
\begin{equation}
\nu =  2L + \sum_i \Delta_i 
\end{equation}
The power $\nu$ is bounded from below; e.g., 
for $A=2$,
$\nu \geq 0$. 
This fact is crucial for the power expansion to be of any use.

\subsection{The Hierarchy of Nuclear Forces}
\begin{figure}[t]
\vspace*{-0.25cm}
\hspace*{1cm}
%\scalebox{0.45}{\includegraphics{diagBW.ps}}
\scalebox{0.45}{\includegraphics{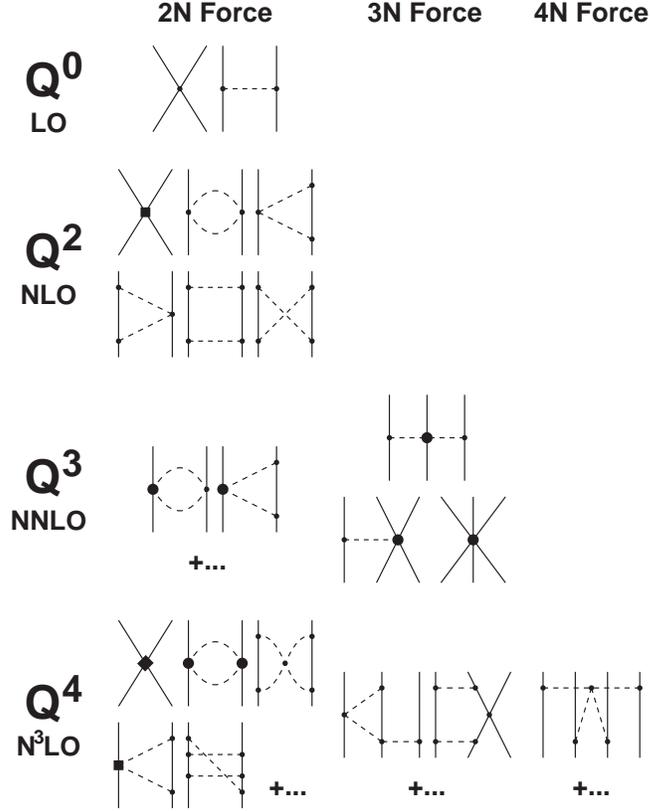}}
\vspace*{-1.00cm}
\caption{Hierarchy of nuclear forces in ChPT. Solid lines
represent nucleons and dashed lines pions. Further explanations are
given in the text.
\label{fig_hi}
}
\end{figure}

Chiral perturbation theory and power counting
imply that nuclear forces emerge as a hierarchy
ruled by the power $\nu$, Fig.~\ref{fig_hi}.

The $NN$ amplitude is determined
by two classes of contributions: contact terms and pion-exchange
diagrams. There are two contacts of order $Q^0$
[${\cal O}(Q^0)$] represented by the four-nucleon graph
with a small-dot vertex shown in the first row of Fig.~\ref{fig_hi}.
The corresponding graph in the second row, four nucleon legs
and a solid square, represents the 
seven contact terms of ${\cal O}(Q^2)$. 
Finally, at ${\cal O}(Q^4)$, we have 15 contact contributions
represented by a four-nucleon graph with a solid diamond.

Now, turning to the pion contributions:
At leading order [LO, ${\cal O}(Q^0)$, $\nu=0$], 
there is only the well-known static one-pion exchange (1PE), second
diagram in the first row of Fig.~\ref{fig_hi}.
Two-pion exchange (2PE) starts
at 
next-to-leading order
 (NLO, $\nu=2$) and all diagrams
of this leading-order two-pion exchange are shown.
Further 2PE contributions occur in any higher order.
Of this sub-leading 2PE, we show only
two representative diagrams at 
next-to-next-to-leading order
(NNLO) and three diagrams at 
next-to-next-to-next-to-leading order
(N$^3$LO).

Finally, there is also three-pion exchange, which
shows up for the first time 
at N$^3$LO (two loops; one representative $3\pi$ diagram
is included in Fig.~\ref{fig_hi}).
At this order, the 3$\pi$ contribution is 
negligible~\cite{Kai99}.

One important advantage of ChPT is that it makes specific
predictions also for many-body forces. For a given order of ChPT,
two-nucleon forces (2NF), three-nucleon forces (3NF), \ldots 
are generated on the same footing 
(cf.\ Fig.~\ref{fig_hi}). 
At LO, there are no 3NF, and
at NLO,
all 3NF terms cancel~\cite{Wei90,Kol94}. 
However, at NNLO and higher orders, well-defined, 
nonvanishing 3NF occur~\cite{Kol94,Epe02b}.
Since 3NF show up for the first time at NNLO, they are weak.
Four-nucleon forces (4NF) occur first at N$^3$LO and, therefore,
they are even weaker.

\section{Two-Nucleon Forces}

In this section, we will elaborate in detail on the
two-nucleon force contributions of which we have given a
rough overview in the previous section.

\subsection{Pion-Exchange Contributions in ChPT \label{sec_pi}}

The effective pion Lagrangians presented in Sec.~\ref{sec_Lpi}
are the crucial ingredients for the evaluation of
the pion-exchange contributions to the $NN$ 
interaction. We will derive these contributions now order
by order.

We will state our results in terms of contributions to the 
momentum-space $NN$ amplitude
in the center-of-mass system (CMS), 
which takes the general form
\begin{eqnarray} 
V({\vec p}~', \vec p) &  = &
%\frac{1}{(2\pi)^3} \;
% \bigg\{ 
% &&
 \:\, V_C \:\, + 
\bbox{\tau}_1 \cdot \bbox{\tau}_2 
\, W_C 
\nonumber \\ &+&  
\left[ \, V_S \:\, + \bbox{\tau}_1 \cdot \bbox{\tau}_2 \, W_S 
\,\:\, \right] \,
\vec\sigma_1 \cdot \vec \sigma_2
\nonumber \\ &+& 
\left[ \, V_{LS} + \bbox{\tau}_1 \cdot \bbox{\tau}_2 \, W_{LS}    
\right] \,
\left(-i \vec S \cdot (\vec q \times \vec k) \,\right)
\nonumber \\ &+& 
\left[ \, V_T \:\,     + \bbox{\tau}_1 \cdot \bbox{\tau}_2 \, W_T 
\,\:\, \right] \,
\vec \sigma_1 \cdot \vec q \,\, \vec \sigma_2 \cdot \vec q  
\nonumber \\ &+& 
\left[ \, V_{\sigma L} + \bbox{\tau}_1 \cdot \bbox{\tau}_2 \, 
      W_{\sigma L} \, \right] \,
\vec\sigma_1\cdot(\vec q\times \vec k\,) \,\,
\vec \sigma_2 \cdot(\vec q\times \vec k\,)
%\bigg\}
\, ,
%\nonumber \\ && 
\label{eq_nnamp}
\end{eqnarray}
where ${\vec p}~'$ and $\vec p$ denote the 
final and initial nucleon momenta 
in the CMS, 
respectively; moreover,
\begin{equation}
\begin{array}{llll}
\vec q &\equiv& {\vec p}~' - \vec p &  \mbox{\rm is the 
momentum transfer},\\
\vec k &\equiv& \frac12 ({\vec p}~' + \vec p) & \mbox{\rm the 
average momentum},\\
\vec S &\equiv& \frac12 (\vec\sigma_1+\vec\sigma_2) & 
\mbox{\rm the total spin},
\end{array}
\label{eq_defqk}
\end{equation}
and $\vec \sigma_{1,2}$ and $\bbox{\tau}_{1,2}$ are 
the spin and isospin 
operators, respectively, of nucleon 1 and 2.
For on-energy-shell scattering, $V_\alpha$ and $W_\alpha$ 
($\alpha=C,S,LS,T,\sigma L$) can be expressed as functions of 
$q$ and $k$ (with
$q\equiv |\vec q|$ and $k\equiv |\vec k|$), only.

Our formalism is similar to the one used by the Munich 
group~\cite{KBW97,Kai01a,Kai01b} except for two
differences: all our momentum space amplitudes
differ by an over-all factor of $(-1)$ and our spin-orbit potentials,
$V_{LS}$ and $W_{LS}$, differ by an additional factor of $(-2)$.
Our conventions are more
in tune with what is commonly used in nuclear physics.

In all expressions given below, we will state only 
the {\it nonpolynomial} contributions to the $NN$ amplitude.
Note, however, that dimensional regularization typically 
generates also polynomial terms.
These polynomials are absorbed by the contact interactions 
to be discussed in a later section and,
therefore, they are of no interest here.

\subsubsection{Zeroth Order (LO) \label{sec_LO}}

At order zero 
[$\nu=0$, ${\cal O}(Q^0)$, lowest order,
leading order, LO], 
there is only the well-known static one-pion exchange, second
diagram in the first row of Fig.~\ref{fig_hi} which is given by:
\begin{equation}
V_{1\pi} ({\vec p}~', \vec p) = - 
%\frac{1}{(2\pi)^3} 
\frac{g_A^2}{4f_\pi^2}
\: 
\bbox{\tau}_1 \cdot \bbox{\tau}_2 
\:
\frac{
\vec \sigma_1 \cdot \vec q \,\, \vec \sigma_2 \cdot \vec q}
{q^2 + m_\pi^2} 
\,.
\label{eq_1pi}
\end{equation}

At first order
[$\nu=1$, ${\cal O}(Q)$], 
there are no pion-exchange contributions
(and also no contact terms).

\begin{figure}[t]
\vspace{-1.00cm}
\hspace{0.50cm}
\scalebox{0.45}{\includegraphics{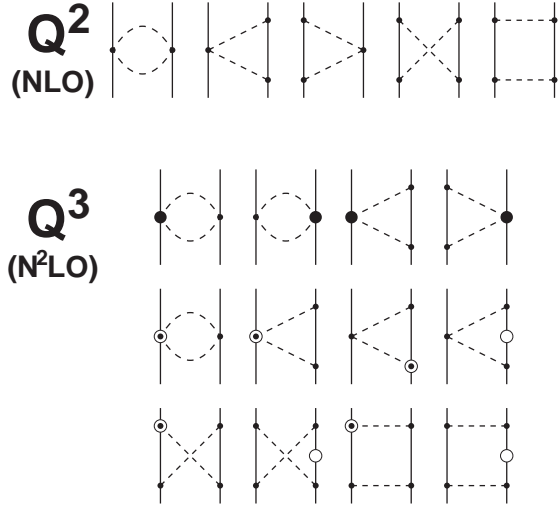}}
\vspace{-4.40cm}
\caption{\small Two-pion exchange contributions to the $NN$
interaction at order two and three in small momenta.
Solid lines represent nucleons and dashed lines pions.
Small dots denote vertices from the 
leading order $\pi N$ Lagrangian
$\widehat{\cal L}^{(1)}_{\pi N}$,
Eq.~(\protect\ref{eq_L1}). 
Large solid dots are vertices
proportional to the LECs $c_i$
from the second order Lagrangian
$\widehat{\cal L}^{(2)}_{\pi N, \, \rm ct}$,
Eq.~(\protect\ref{eq_L2ct}).
Symbols with an open circles are relativistic $1/M_N$ corrections
contained in the second order Lagrangian
$\widehat{\cal L}^{(2)}_{\pi N}$,
Eqs.~(\protect\ref{eq_L2}).
Only a few representative examples of $1/M_N$ corrections are
shown and not all.
\label{fig_diag1}
}
\end{figure}

\subsubsection{Second Order (NLO) \label{sec_NLO}}

Non-vanishing higher-order graphs start at second order
($\nu=2$, next-to-leading order, NLO). 
The most efficient way to evaluate these loop diagrams
is to
use covariant perturbation theory and dimensional
regularization. 
This is the method applied by the Munich 
group~\cite{KBW97,Kai01a,Kai01b}. 
One starts
with the relativistic versions of the $\pi N$ Lagrangians
(cf.\ Sec.~\ref{sec_Lpi}) and sets up four-dimensional
(covariant) loop integrals.
Relativistic vertices and nucleon propagators are then 
expanded in powers of $1/M_N$. 
The divergences that occur in conjunction with the four-dimensional
loop integrals are treated by means of dimensional regularization,
a prescription which is
consistent with chiral symmetry and power counting.
The results derived in this way
are the same obtained when starting right away
with the HB versions of the $\pi N$ Lagrangians.
However, as it turns out, the method used by
the Munich group is more efficient in dealing 
with the rather tedious calculations.

Two-pion exchange occurs first at second order, 
also know as leading-order $2\pi$ exchange. 
The graphs are shown in the first row of 
Fig.~\ref{fig_diag1}.
Since a loop creates already $\nu=2$,
the vertices involved at this order can only be from the leading/lowest order
Lagrangian $\widehat{\cal L}^{(1)}_{\pi N}$, Eq.~(\ref{eq_L1}),
i.~e., they carry only one derivative.
These vertices are denoted by small dots in
Fig.~\ref{fig_diag1}.
Concerning the box diagram, we should note that
we include only the non-iterative part
of this diagram which is obtained by subtracting 
the iterated 1PE contribution 
Eq.~(\ref{eq_2piitKBW}) or
Eq.~(\ref{eq_2piitEM}), 
below, but using
$M_N^2/E_p 
\approx M_N^2/E_{p''} 
\approx M_N$
at this order (NLO).
Summarizing all contributions from irreducible two-pion exchange
at second order, one obtains~\cite{KBW97}:
\begin{eqnarray} 
W_C &=&-{L(q)\over384\pi^2 f_\pi^4} 
\left[4m_\pi^2(5g_A^4-4g_A^2-1)
+q^2(23g_A^4 -10g_A^2-1) 
\right.
\nonumber \\ &&
\left.
+ {48g_A^4 m_\pi^4 \over w^2} \right] \,,  
\label{eq_2C}
\\   
V_T &=& -{1\over q^2} V_{S} 
    \; = \; -{3g_A^4 L(q)\over 64\pi^2 f_\pi^4} \,, 
\label{eq_2T}
\end{eqnarray}  
where
\begin{equation} 
L(q)  \equiv  {w\over q} \ln {w+q \over 2m_\pi}
\end{equation}
and
\begin{equation} 
 w  \equiv  \sqrt{4m_\pi^2+q^2} \,. 
\end{equation}

\subsubsection{Third Order (NNLO) \label{sec_NNLO}}

The two-pion exchange diagrams of order three ($\nu=3$, 
next-to-next-to-leading order,
NNLO) are very similar to the ones of order two, except that 
they contain one insertion from 
$\widehat{\cal L}^{(2)}_{\pi N}$, Eq.~(\ref{eq_L2}). 
The resulting contributions are typically either 
proportional to one of the 
low-energy constants $c_i$ or they contain a factor $1/M_N$.
Notice that relativistic $1/M_N$ corrections can occur
for vertices and nucleon propagators.
In Fig.~\ref{fig_diag1}, we show in row 2 the diagrams with
vertices proportional to $c_i$ (large solid dot), 
Eq.~(\ref{eq_L2ct}),
and in row 3 and 4 a few representative graphs with a $1/M_N$ 
correction (symbols with an open circle). The number 
of $1/M_N$ correction graphs
is large and not all are shown in the figure.
Again, the box diagram is corrected for a contribution from
the iterated 1PE. If the iterative 2PE of 
Eq.~(\ref{eq_2piitKBW}) is used, the expansion of
the factor 
$M^2_N/E_p = M_N - p^2/2M_N + \ldots$ 
is applied and
the term proportional to $(-p^2/2M_N)$ is subtracted from
the third order box diagram contribution.
Then, one obtains for the full third order contribution~\cite{KBW97}:
\begin{eqnarray} 
V_C &=&{3g_A^2 \over 16\pi f_\pi^4} 
\left\{ 
 {g_A^2 m_\pi^5  \over 16M_N w^2}  
-\left[2m_\pi^2( 2c_1-c_3)-q^2  \left(c_3 +{3g_A^2\over16M_N}\right)
\right]
\right.
\nonumber \\ && \times
\left.
\widetilde{w}^2 A(q) \right\} \,, 
\label{eq_3C}
\\
W_C &=& {g_A^2\over128\pi M_N f_\pi^4} \left\{ 
 3g_A^2 m_\pi^5 w^{-2} 
\right.
\nonumber \\ &&
\left.
 - \left[ 4m_\pi^2 +2q^2-g_A^2(4m_\pi^2+3q^2) \right] 
\widetilde{w}^2 A(q)
\right\} 
\,,\\ 
V_T &=& -{1 \over q^2} V_{S}
   \; = \; {9g_A^4 \widetilde{w}^2 A(q) \over 512\pi M_N f_\pi^4} 
 \,,  \\ 
W_T &=&-{1\over q^2}W_{S} 
\nonumber \\
    &=&-{g_A^2 A(q) \over 32\pi f_\pi^4}
\left[
\left( c_4 +{1\over 4M_N} \right) w^2
-{g_A^2 \over 8M_N} (10m_\pi^2+3q^2)  \right] 
\,,
\label{eq_3T}
\\
V_{LS} &=&  {3g_A^4  \widetilde{w}^2 A(q) \over 32\pi M_N f_\pi^4} 
 \,,\\  
W_{LS} &=& {g_A^2(1-g_A^2)\over 32\pi M_N f_\pi^4} 
w^2 A(q) \,, 
\label{eq_3LS}
\end{eqnarray}   
with
\begin{equation} 
A(q) \equiv {1\over 2q}\arctan{q \over 2m_\pi} 
\end{equation}
and
\begin{equation} 
\widetilde{w} \equiv  \sqrt{2m_\pi^2+q^2} \,. 
\end{equation}

As discussed in Sec.~\ref{sec_2piit}, below,
we prefer the iterative 2PE defined in Eq.~(\ref{eq_2piitEM}),
which leads to a different NNLO term for the iterative 2PE.
This changes the $1/M_N$ terms
in the above potentials. The changes are obtained
by adding to Eqs.~(\ref{eq_3C})-(\ref{eq_3T}) the
following terms:

\begin{eqnarray}
V_C &=& -\frac{3 g_A^4}{256 \pi f_\pi^4 M_N} 
(m_\pi \omega^2 + \tilde \omega^4 A(q) )
\label{eq_3EM1}
\\
W_C &=& \frac{g_A^4}{128 \pi f_\pi^4 M_N} 
(m_\pi \omega^2 + \tilde \omega^4 A(q) )
\\
V_T &=& -\frac{1}{q^2} V_S = \frac{3 g_A^4}{512 \pi f_\pi^4 M_N} 
(m_\pi + \omega^2 A(q) )
\\
W_T &=& -\frac{1}{q^2} W_S = -\frac{g_A^4}{256 \pi f_\pi^4 M_N} 
(m_\pi + \omega^2 A(q) )
\label{eq_3EM4}
\end{eqnarray}

\begin{figure}[t]
%\vspace{-1cm}
\hspace{1.0cm}
\scalebox{0.45}{\includegraphics{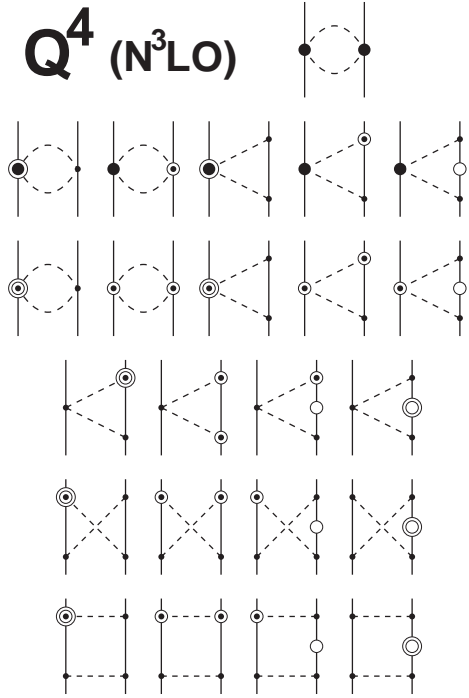}}
\vspace{-3.25cm}
\caption{\small One-loop $2\pi$-exchange contributions to the $NN$
interaction at order four. Basic notation as in Fig.~\ref{fig_diag1}.
Symbols with a large solid dot and an open circle denote 
$1/M_N$ corrections of vertices
proportional to $c_i$.
Symbols with two open circles mark 
relativistic $1/M^2_N$ corrections.
Both corrections are part of the third order Lagrangian
$\widehat{\cal L}^{(3)}_{\pi N}$, Eq.~(\ref{eq_L3}).
Representative examples for all types of one-loop graphs that occur 
at this order are shown.
\label{fig_diag2}}
\end{figure}

\begin{figure}[t]
%\vspace{-1.0cm}
\hspace{1.0cm}
\scalebox{0.50}{\includegraphics{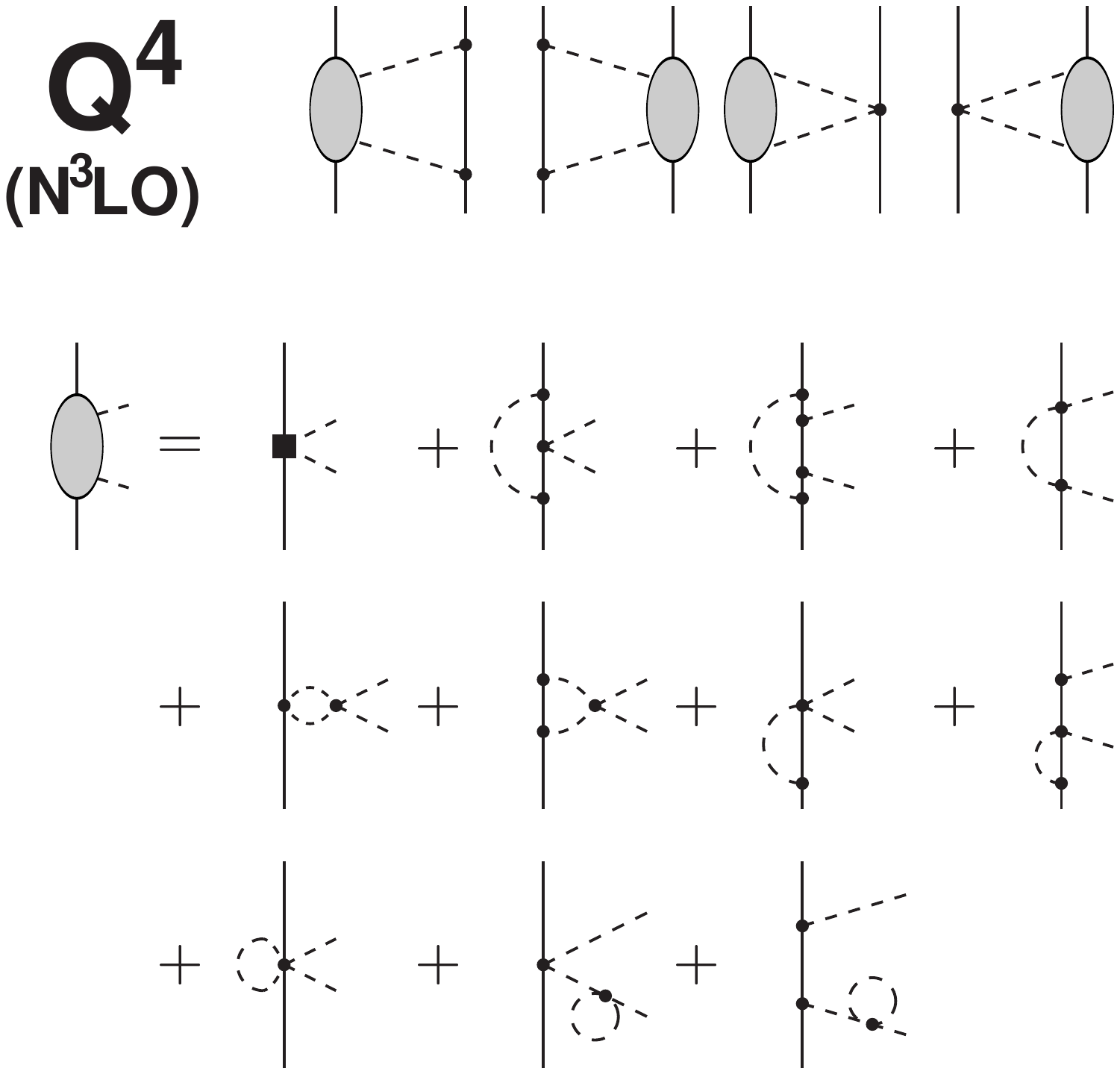}}
\vspace{-6.50cm}
\caption{\small Two-loop $2\pi$-exchange contributions at order four.
Basic notation as in Fig.~\ref{fig_diag1}.
The oval stands for all one-loop $\pi N$ graphs
some of which are shown in the lower part of the figure.
The solid square represents vertices proportional to the LECs
$d_i$ which are introduced by the third order Lagrangian
${\cal L}^{(3)}_{\pi N}$,
Eq.~(\ref{eq_L3rel}). More explanations are given in the text.
\label{fig_diag3}}
\end{figure}

\subsubsection{Fourth Order (N$^3$LO) \label{sec_N3LO}}

This order, which may also be denoted by
next-to-next-to-next-to-leading order (N$^3$LO), is 
very involved.
Three-pion exchange (3PE) occurs for the first time
at this order. The 3PE contribution at N$^3$LO has been
calculated by the Munich group and found to be
negligible~\cite{Kai99}. Therefore, we will ignore it.

The 2PE contributions at N$^3$LO
can be subdivided into two groups,
one-loop graphs, 
Fig.~\ref{fig_diag2}, 
and
two-loop diagrams, 
Fig.~\ref{fig_diag3}.
Since these contributions are very complicated,
we have moved them to Appendix A.

\subsubsection{Iterated One-Pion-Exchange \label{sec_2piit}}

Besides all the irreducible 2PE contributions presented above,
there is also the reducible 2PE which is generated
from iterated 1PE. This 
``iterative 2PE'' is the only 2PE contribution
which produces an imaginary part. Thus, one wishes
to formulate this contribution such that relativistic
elastic unitarity is satisfied.
There are several ways to achieve this.

Kaiser {\it et al.}~\cite{KBW97} define the iterative
2PE contribution as follows,
\begin{equation}
V_{2\pi, it}^{\rm(KBW)} ({\vec p}~',{\vec p})  = 
\:
\frac{M_N^2}{E_{p}} 
\:
\int 
\frac{d^3p''}{(2\pi)^3} 
\:
\frac{V_{1\pi}({\vec p}~',{\vec p}~'')\,
V_{1\pi}({\vec p}~'',{\vec p})} 
{{ p}^{2}-{p''}^{2}+i\epsilon}
\label{eq_2piitKBW}
\end{equation}
with $V_{1\pi}$ given in Eq.~(\ref{eq_1pi}).

Since we adopt the relativistic scheme developed
by Blankenbecler and Sugar~\cite{BS66} (BbS)
(see beginning of Sec.~\ref{sec_pot}),
we prefer the following formulation which is
consistent with the BbS approach (and, of course,
with relativistic elastic unitarity): 
\begin{equation}
V_{2\pi, it}^{\rm(EM)} ({\vec p}~',{\vec p})  = 
\int 
\frac{d^3p''}{(2\pi)^3} 
\:
\frac{M_N^2}{E_{p''}} 
\:
\frac{V_{1\pi}({\vec p}~',{\vec p}~'')\,
V_{1\pi}({\vec p}~'',{\vec p})} 
{{ p}^{2}-{p''}^{2}+i\epsilon}
\,.
\label{eq_2piitEM}
\end{equation}

The iterative 2PE contribution has to be subtracted
from the covariant box diagram, order by order.
For this, the expansion
$M^2_N/E_p = M_N - p^2/2M_N + \ldots$ 
is applied in Eq.~(\ref{eq_2piitKBW}) and
$M^2_N/E_{p''} = M_N - {p''}^2/2M_N + \ldots$ 
in Eq.~(\ref{eq_2piitEM}).
At NLO, both choices for the iterative 2PE
collapse to the same, while at NNLO there are 
obvious differences.

\subsection{$NN$ Scattering in Peripheral Partial Waves
Using the Perturbative Amplitude \label{sec_peri}}

After the tedious mathematics of the previous section,
it is time for more tangible affairs.
The obvious question to address now is:
How does the derived $NN$ amplitude compare to 
empirical information?
Since our derivation includes only one- and two-pion exchanges,
we are dealing here with the long- and intermediate-range
part of the $NN$ interaction. This part of the nuclear force
is probed in the peripheral partial waves of $NN$ scattering. 
Thus, in this section, we will calculate the phase shifts
that result from the $NN$ amplitudes presented in the previous
section and compare them to the empirical phase shifts
as well as to the predictions from conventional meson theory. 
Besides the irreducible two-pion exchanges derived
above, we must also include
1PE and iterated 1PE.

In this section~\cite{note3}, which is restricted to just peripheral
waves, we will always consider neutron-proton ($np$) scattering 
and take the charge-dependence of 1PE due to pion-mass splitting
into account, since it is appreciable. With the definition
\begin{equation}
V_{1\pi} (m_\pi) \equiv -\, 
%\frac{1}{(2\pi)^3} 
\frac{g_A^2}{4f_\pi^2}\, \frac{
\vec \sigma_1 \cdot \vec q \,\, \vec \sigma_2 \cdot \vec q}
{q^2 + m_\pi^2} 
\,,
\end{equation}
the charge-dependent 1PE for $np$ scattering is
\begin{equation}
V_{1\pi}^{(np)} ({\vec p}~', \vec p) 
= -V_\pi(m_{\pi^0}) + (-1)^{I+1}\, 2\, V_\pi (m_{\pi^\pm})
\,,
\label{eq_1PE}
\end{equation}
where $I$ denotes the isospin of the two-nucleon system.
We use $m_{\pi^0}=134.9766$ MeV,
 $m_{\pi^\pm}=139.5702$ MeV~\cite{PDG},
and
\begin{equation}
M_N  =  \frac{2M_pM_n}{M_p+M_n} = 938.9182 \mbox{ MeV}
\,.
\end{equation}
Also in the iterative 2PE, we apply the charge-dependent
1PE, i.e., in Eq.~(\ref{eq_2piitEM}) we replace $V_{1\pi}$
with $V_{1\pi}^{(np)}$.

The perturbative relativistic T-matrix for $np$ scattering
in peripheral waves is 
\begin{equation}
T({\vec p}~',\vec p) = V_{1\pi}^{(np)} ({\vec p}~', \vec p) 
+V_{2\pi, it}^{({\rm EM},np)} ({\vec p}~',{\vec p}) +V_{2\pi, irr} 
({\vec p}~',{\vec p}) 
\,,
\label{eq_tall}
\end{equation}
where $V_{2\pi,irr}$ refers to any or all
of the irreducible 2PE contributions presented 
in Sec.~\ref{sec_pi},
depending on the order at which the calculation is conducted. 
In the calculation of the irreducible 2PE,
we use the average pion mass $m_\pi = 138.039$ MeV and, thus,
neglect the charge-dependence due to pion-mass splitting.
The charge-dependence that emerges from irreducible $2\pi$ 
exchange was investigated in Ref.~\cite{LM98} and found to be
negligible for partial waves with $L\geq 3$.

\begin{table}[t]
\caption{Low-energy constants, LECs, used for a $NN$ potential
at N$^3$LO, Sec.~\ref{sec_potn3lo}, and in the calculation of 
the peripheral $NN$ phase shifts shown in 
Fig.~\ref{fig_f} (column 
``$NN$ periph.\ Fig.~\ref{fig_f}''). 
The $c_i$ belong to the dimension-two $\pi N$ Lagrangian,
Eq.~(\ref{eq_L2ct}), and are in units of GeV$^{-1}$,
while the $\bar{d}_i$ are associated with the dimension-three
Lagrangian, Eq.~(\ref{eq_L3rel}), and
 are in units of GeV$^{-2}$.
The column ``$\pi N$ empirical'' shows determinations from $\pi N$
data.
\label{tab_LEC}}
\smallskip
\begin{tabular*}{\textwidth}{@{\extracolsep{\fill}}cccc}
\hline
\hline
\noalign{\smallskip}
 LEC & $NN$ potential & $NN$ periph.      & $\pi N$ \\
     & at N$^3$LO     &  Fig.~\ref{fig_f} & empirical \\
\hline
\noalign{\smallskip}
$c_1$ & --0.81 & --0.81 & $-0.81\pm 0.15^a$ \\
$c_2$ & 2.80 & 3.28 & $3.28\pm 0.23^b$ \\
$c_3$ & --3.20 & --3.40 & $-4.69\pm 1.34^a$ \\
$c_4$ & 5.40 & 3.40 & $3.40\pm 0.04^a$ \\
$\bar{d}_1 + \bar{d}_2$ & 3.06 & 3.06 & $3.06\pm 0.21^b$ \\
$\bar{d}_3$ & --3.27 & --3.27 & $-3.27\pm 0.73^b$ \\
$\bar{d}_5$ & 0.45 & 0.45 & $0.45\pm 0.42^b$ \\
$\bar{d}_{14} - \bar{d}_{15}$ & --5.65 & --5.65 & $-5.65\pm 0.41^b$ \\
\hline
\hline
\end{tabular*}
$^a$Table~1, Fit~1 of Ref.~\cite{BM00}.\\
$^b$Table~2, Fit~1 of Ref.~\cite{FMS98}.
\end{table}

For the $T$-matrix given in Eq.~(\ref{eq_tall}),
we calculate phase shifts for
partial waves with $L\geq 3$ and $T_{lab}\leq 300$ MeV.
At order four in small momenta, partial waves
with $L\geq 3$ do not receive any contributions from
contact interactions and, thus, the non-polynomial
pion contributions uniquely predict the $F$ and
higher partial waves.
We use $f_\pi = 92.4$ MeV~\cite{PDG}
and $g_A = 1.29$. 
Via the Goldberger-Treiman relation,
$g_A  =  g_{\pi NN} \; f_\pi/M_N$, 
our value for $g_A$ is consistent
with $g_{\pi NN}^2/4\pi = 13.63\pm 0.20$
which is obtained from $\pi N$ and $NN$ analysis~\cite{STS93,AWP94}.

\begin{figure}[t]
\vspace*{-2cm}
\hspace*{0.5cm}
\scalebox{0.50}{\includegraphics{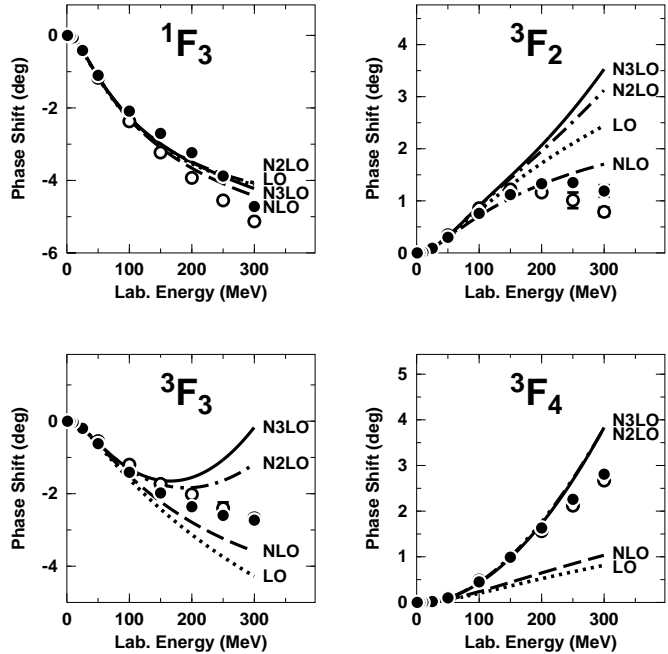}}
\vspace*{-2.5cm}
\caption{\small $F$-wave phase shifts of neutron-proton scattering
for laboratory kinetic energies below 300 MeV.
We show the predictions from chiral pion exchange
at leading order (LO), next-to-leading order (NLO),
next-to-next-to-leading order (N2LO), and
next-to-next-to-next-to-leading order (N3LO).
The solid dots and open circles are the results from the Nijmegen
multi-energy $np$ phase shift analysis~\protect\cite{Sto93} 
and the VPI
single-energy $np$ analysis SM99~\protect\cite{SM99}, respectively.
\label{fig_f}}
\end{figure}

The LECs used in this calculation are shown in Table~\ref{tab_LEC},
column ``$NN$ periph.\ Fig.~\ref{fig_f}''.
Note that many determinations of the LECs, $c_i$ and $\bar{d}_i$,
can be found in the literature.
The most reliable way to determine the LECs from empirical
$\pi N$ information is to extract them from the $\pi N$ amplitude
inside the Mandelstam triangle (unphysical region) which can
be constructed with the help of dispersion relations from empirical
$\pi N$ data. This method was used by B\"uttiker and Mei\ss ner~\cite{BM00}.
Unfortunately, the values for $c_2$ and all $\bar{d}_i$ parameters
obtained in Ref.~\cite{BM00} carry uncertainties,
so large that the values cannot provide any guidance.
Therefore, in Table~\ref{tab_LEC}, only  $c_1$, $c_3$, and $c_4$
are from Ref.~\cite{BM00}, while the other LECs
are taken from Ref.~\cite{FMS98} where the $\pi N$ amplitude in the
physical region was considered.
To establish a link between $\pi N$ and $NN$, we apply 
the values from the above determinations in our calculations
of the $NN$ peripheral phase shifts.
In general, we use the mean values;
the only exception is $c_3$, where we choose
a value that is, in terms of magnitude, about one standard
deviation below the one from Ref.~\cite{BM00}.
With the exception of $c_3$,
phase shift predictions do not depend sensitively on
variations of the LECs within the quoted uncertainties.

In Fig.~\ref{fig_f}, 
we show the phase-shift
predictions for neutron-proton scattering in $F$
waves for laboratory kinetic energies below 300 MeV
(for $G$ and $H$ waves, see Ref.~\cite{EM02}).
The orders displayed are defined as follows:
\begin{itemize}
\item
Leading order (LO) is just 1PE, 
Eq.~(\ref{eq_1PE}).
\item
Next-to-leading order (NLO) is 1PE,
Eq.~(\ref{eq_1PE}),
 plus iterated 1PE,
Eq.~(\ref{eq_2piitEM}), plus the contributions of Sec.~\ref{sec_NLO}
(order two), Eqs.~(\ref{eq_2C}) and (\ref{eq_2T}).
\item
Next-to-next-to-leading order (denoted by N2LO in the figures)
consists of NLO plus the contributions of Sec.~\ref{sec_NNLO} (order three),
Eqs.~(\ref{eq_3C})-(\ref{eq_3LS}) and
(\ref{eq_3EM1})-(\ref{eq_3EM4}).
\item
Next-to-next-to-next-to-leading order (denoted by N3LO in the figures)
consists of N2LO plus the contributions of Sec.~\ref{sec_N3LO}
(order four),
Eqs.~(\ref{eq_4c2C})-(\ref{eq_4M2sL}) 
and (\ref{eq_42lC})-(\ref{eq_42lT}). 
\end{itemize}
It is clearly seen in 
Fig.~\ref{fig_f} 
that the leading order $2\pi$ exchange (NLO)
is a rather small contribution, insufficient to explain
the empirical facts. In contrast, the next order (N2LO)
is very large, several times NLO. This is due to the
$\pi\pi N N$ contact interactions proportional
to the LECs $c_i$ that are introduced by the
second order Lagrangian 
${\cal L}^{(2)}_{\pi N}$, Eq.~(\ref{eq_L2rel}). 
These contacts are supposed to simulate the contributions
from intermediate $\Delta$-isobars and correlated $2\pi$
exchange which are known
to be large (see, e.~g., Ref.~\cite{MHE87}).

\begin{figure}[t]
\vspace*{-2cm}
\hspace*{0.5cm}
\scalebox{0.50}{\includegraphics{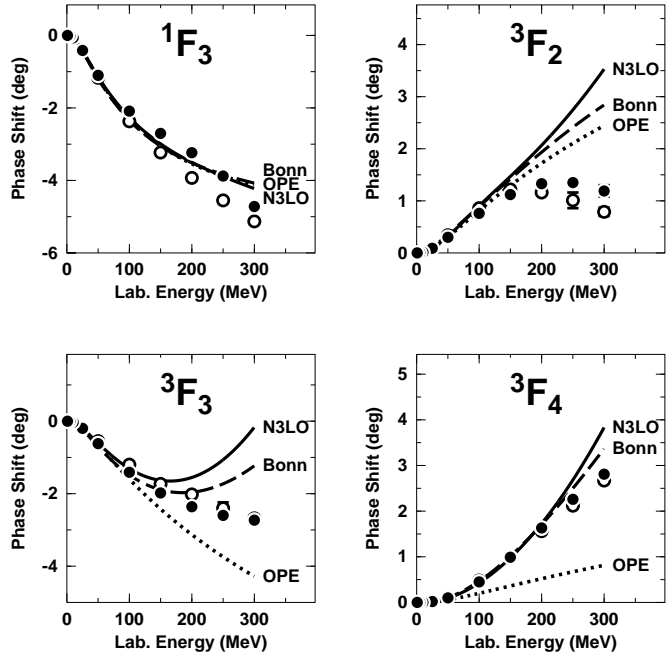}}
\vspace*{-2.5cm}
\caption{\small $F$-wave phase shifts of neutron-proton scattering
for laboratory kinetic energies below 300 MeV.
We show the results from one-pion-exchange (OPE),
and one- plus two-pion exchange as predicted by ChPT at
next-to-next-to-next-to-leading order (N3LO) and 
by the Bonn Full Model~\protect\cite{MHE87} (Bonn).
Note that the ``Bonn'' curve does not include the repulsive
$\omega$ and $\pi\rho$ exchanges of the full model, since 
this figure serves the purpose to compare just predictions
by different models/theories for the $\pi+2\pi$ contribution 
to the $NN$ interaction.
Empirical phase shifts (solid dots and open circles)
as in Fig.~\protect\ref{fig_f}.
\label{fig_ff}}
\end{figure}

At N3LO a clearly identifiable
trend towards convergence emerges.
Obviously, $^1F_3$ and $^3F_4$ appear fully converged.
However, in $^3F_2$ and $^3F_3$, N3LO differs
noticeably from NNLO, but the difference is much smaller than the
one between NNLO and NLO. This is what we perceive as a trend towards
convergence.

In Fig.~\ref{fig_ff}, 
we conduct a comparison between the predictions from chiral one- and
two-pion exchange at N3LO and the corresponding predictions from conventional
meson theory (curve `Bonn'). 
As representative for conventional meson theory, we choose the Bonn
meson-exchange model for the $NN$ interaction~\cite{MHE87},
since it contains a comprehensive and thoughtfully constructed
model for $2\pi$ exchange.
This $2\pi$ model includes box and crossed box diagrams with
$NN$, $N\Delta$, and $\Delta\Delta$ intermediate states
as well as direct $\pi\pi$ interaction in $S$- and $P$-waves
(of the $\pi\pi$ system)
consistent with empirical information from $\pi N$ and $\pi\pi$
scattering.
Besides this the Bonn model also includes (repulsive) $\omega$-meson
exchange and irreducible diagrams of $\pi$ and $\rho$ exchange
(which are also repulsive).
However, note that
in the phase shift predictions displayed in Fig.~\ref{fig_ff}, 
the ``Bonn'' curve includes only the $1\pi$ and $2\pi$ contributions
from the Bonn model; the short-range contributions are
left out since the purpose of the figure is to compare
different models/theories for $\pi+2\pi$.
In all waves shown 
we see, in general, good agreement between N3LO and Bonn.
In $^3F_2$ and $^3F_3$ above 150 MeV and in
$^3F_4$ above 250 MeV the chiral model at N3LO is more
attractive than the Bonn $2\pi$ model. Note, however,
that the Bonn model is relativistic and, thus, includes
relativistic corrections up to infinite orders.
Thus, one may speculate that higher orders in 
ChPT may create some
repulsion, moving the Bonn and the chiral predictions even closer
together~\cite{foot2}.

The $2\pi$ exchange contribution to the $NN$ interaction
can also be derived
from {\it empirical} $\pi N$ and $\pi\pi$ input
using dispersion theory, which is based upon unitarity,
causality (analyticity), and crossing symmetry.
The amplitude $N\bar{N}\rightarrow \pi\pi$ is
constructed from $\pi N \rightarrow \pi N$
and $\pi N \rightarrow \pi\pi N$ data using 
crossing properties and analytic
continuation; this amplitude is then `squared' to yield 
the $N\bar{N}$ amplitude 
which is related to $NN$ by crossing symmetry~\cite{BJ76}.
The Paris group~\cite{Vin79,Lac80} pursued this path and 
calculated $NN$ phase shifts in peripheral partial waves.
Naively, the dispersion-theoretic approach is
the ideal one, since it is based exclusively on empirical
information. Unfortunately, in practice,
quite a few uncertainties enter into the approach.
First, there are ambiguities
in the analytic continuation and, second, 
the dispersion integrals have to be cut off at a certain
momentum to ensure reasonable results.
In Ref.~\cite{MHE87}, a thorough comparison was conducted
between the predictions by the Bonn model and the
Paris approach and it was demonstrated that the Bonn
predictions always lie comfortably within the range of uncertainty
of the dispersion-theoretic results.
Therefore, there is no need to perform a
separate comparison of our chiral N3LO predictions with
dispersion theory, since it would not add anything that we
cannot conclude from Fig.~\ref{fig_ff}.

Finally, we need to compare the
predictions with the empirical phase shifts.
In $F$ waves the N$^3$LO predictions above 200 MeV are,
in general, too attractive. Note, however, that this is
also true for the predictions by the Bonn $\pi + 2\pi$
model. In the {\it full} Bonn model, 
besides $\pi+2\pi$, 
(repulsive)
$\omega$ and $\pi\rho$ exchanges are included
which move
the predictions right on top of the data.
The exchange of a $\omega$ meson or  combined $\pi\rho$
exchange are $3\pi$ exchanges.
Three-pion exchange occurs first at chiral order four.
It has be investigated by Kaiser~\cite{Kai99} and found to be
negligible, at this order.
However, $3\pi$ exchange at order five appears to be sizable~\cite{Kai01}
and may have impact on $F$ waves.
Besides this, there is the usual short-range phenomenology.
In ChPT, this short-range interaction
is parametrized in terms of four-nucleon contact terms
(since heavy mesons do not have a place in that theory).
Contact terms of order four (N$^3$LO) do not contribute to
$F$-waves, but order six does.
In summary, the remaining small 
discrepancies between the N3LO predictions and the empirical
phase shifts may be straightened out in
fifth or sixth order of ChPT.

\subsection{$NN$ Contact Potentials \label{sec_ct}}

In conventional meson theory, the short-range nuclear force
is described by the exchange of heavy mesons, notably the
$\omega(782)$. 
The qualitative short-distance behavior of the $NN$ 
potential is obtained
by Fourier transform of the propagator of
a heavy meson,
\begin{equation}
\int d^3q \frac{e^{i{\vec q} \cdot {\vec r}}}{m^2_\omega
+ {\vec q}^2} \;
\sim \;
 \frac{e^{-m_\omega r}}{r} \; .
\end{equation}

ChPT is an expansion in small momenta $Q$, too small
to resolve structures like a $\rho(770)$ or $\omega(782)$
meson, because $Q \ll \Lambda_\chi \approx m_{\rho,\omega}$.
But the latter relation allows us to expand the propagator 
of a heavy meson into a power series,
\begin{equation}
\frac{1}{m^2_\omega + Q^2} 
\approx 
\frac{1}{m^2_\omega} 
\left( 1 
- \frac{Q^2}{m^2_\omega}
+ \frac{Q^4}{m^4_\omega}
-+ \ldots
\right)
,
\end{equation}
where the $\omega$ is representative
for any heavy meson of interest.
The above expansion suggests that it should be 
possible to describe the short distance part of
the nuclear force simply in terms of powers of
$Q/m_\omega$, which fits in well
with our over-all 
power scheme since $Q/m_\omega \approx Q/\Lambda_\chi$.

A second purpose of contact terms is renormalization.
Dimensional regularization
of the loop integrals of pion-exchanges (cf.\
Sec.~\ref{sec_pi}) typically generates polynomial terms
with coefficients that are, in part, infinite or scale
dependent. Contact terms pick up infinities
and remove scale dependence.

The partial-wave decomposition of a power
$Q^\nu$ has an interesting property.
First note that $Q$ can only be either
the momentum transfer between the two interacting
nucleons $q$ or the average momentum $k$
[cf.\ Eq.~(\ref{eq_defqk}) for their definitions].
In any case, for even $\nu$,
\begin{equation}
Q^\nu = 
f_{\frac{\nu}{2}}(\cos \theta) 
\, ,
\end{equation}
where $f_m$ stands for a polynomial of degree $m$
and $\theta$ is the CMS scattering angle.
The partial-wave decomposition of $Q^\nu$ for a state
of orbital-angular momentum $L$
involves the integral
\begin{equation}
I^{(\nu)}_L  
=\int_{-1}^{+1} Q^\nu P_L(\cos \theta) d\cos \theta 
=\int_{-1}^{+1}
f_{\frac{\nu}{2}}(\cos \theta) 
 P_L(\cos \theta) d\cos \theta 
\,,
\end{equation}
where $P_L$ is a Legendre polynomial.
Due to the orthogonality of the $P_L$, 
\begin{equation}
I^{(\nu)}_L = 0  
\hspace*{.5cm}
\mbox{for}
\hspace*{.5cm}
L > \frac{\nu}{2} \, .
\end{equation}
Consequently, contact terms of order zero contribute only
in $S$-waves, while order-two terms contribute up to 
$P$-waves, order-four terms up to $D$-waves,
etc..

We will now present, one by one, the various orders of 
$NN$ contact terms together with their partial-wave
decomposition~\cite{EAH71}.
Note that, due to parity, only even powers of $Q$
are allowed.

\subsubsection{Zeroth Order}

The contact potential at order zero reads:
\begin{equation}
V^{(0)}(\vec{p'},\vec{p}) =
C_S +
C_T \, \vec{\sigma}_1 \cdot \vec{\sigma}_2 
\end{equation}
Partial wave decomposition yields:
\be
V^{(0)}(^1 S_0)          &=&  \widetilde{C}_{^1 S_0} =
4\pi\, ( C_S - 3 \, C_T )
\nonumber \\
V^{(0)}(^3 S_1)          &=&  \widetilde{C}_{^3 S_1} =
4\pi\, ( C_S + C_T )
\ee

\subsubsection{Second Order}

The contact potential contribution of order two is
given by:
\be
V^{(2)}(\vec{p'},\vec{p}) &=&
C_1 q^2 +
C_2 k^2 
\nonumber 
\\ &+& 
\left(
C_3 q^2 +
C_4 k^2 
\right) \vec{\sigma}_1 \cdot \vec{\sigma}_2 
\nonumber 
\\
&+& C_5 \left( -i \vec{S} \cdot (\vec{q} \times \vec{k}) \right)
\nonumber 
\\ &+& 
 C_6 ( \vec{\sigma}_1 \cdot \vec{q} )\,( \vec{\sigma}_2 \cdot 
\vec{q} )
\nonumber 
\\ &+& 
 C_7 ( \vec{\sigma}_1 \cdot \vec{k} )\,( \vec{\sigma}_2 \cdot 
\vec{k} )
\ee
Second order partial wave contributions:
\footnotesize
\be
V^{(2)}(^1 S_0)          &=&  C_{^1 S_0} ( p^2 + {p'}^2 ) 
\nonumber \\
&=&
4\pi\, \left( C_1 + \frac{1}{4} C_2 - 3 C_3 - \frac{3}{4} C_4 - 
C_6 - \frac{1}{4} C_7 \right)
( p^2 + {p'}^2 )
\nonumber 
\\
V^{(2)}(^3 P_0)          &=&  C_{^3 P_0} \, p p'
\nonumber \\
&=&
4\pi\, \left( -\frac{2}{3} C_1 + \frac{1}{6} C_2 - \frac{2}{3} C_3 
+ \frac{1}{6} C_4 - \frac{2}{3} C_5
+ 2 C_6 - \frac{1}{2} C_7 \right)
 p p' 
\nonumber 
\\
V^{(2)}(^1 P_1)          &=&  C_{^1 P_1} \, p p' 
\nonumber \\
&=&
4\pi\, \left( -\frac{2}{3} C_1 + \frac{1}{6} C_2 + 2 C_3 
- \frac{1}{2} C_4 
+ \frac{2}{3} C_6 - \frac{1}{6} C_7 \right)
 p p' 
\nonumber 
\\
V^{(2)}(^3 P_1)          &=&  C_{^3 P_1} \, p p' 
\nonumber \\
&=&
4\pi\, \left( -\frac{2}{3} C_1 + \frac{1}{6} C_2 - \frac{2}{3} C_3 
+ \frac{1}{6} C_4 - \frac{1}{3} C_5
- \frac{4}{3} C_6 + \frac{1}{3} C_7 \right)
 p p' 
\nonumber 
\\
V^{(2)}(^3 S_1)          &=&  C_{^3 S_1} ( p^2 + {p'}^2 ) 
\nonumber \\
&=&
4\pi\, \left( C_1 + \frac{1}{4} C_2 + C_3 + \frac{1}{4} C_4 + 
\frac{1}{3} C_6 + \frac{1}{12} C_7 \right)
( p^2 + {p'}^2 )
\nonumber 
\\
V^{(2)}(^3 S_1- ^3 D_1)  &=&  C_{^3 S_1- ^3 D_1}  p^2 
\nonumber \\
&=&
4\pi\, \left( -\frac{2\sqrt{2}}{3} C_6 - \frac{2\sqrt{2}}{12} 
C_7 \right)
p^2
\nonumber 
\\
V^{(2)}(^3 P_2)          &=&  C_{^3 P_2} \, p p' 
\nonumber \\
&=&
4\pi\, \left( -\frac{2}{3} C_1 + \frac{1}{6} C_2 - \frac{2}{3} C_3 
+ \frac{1}{6} C_4 + \frac{1}{3} C_5 \right)
 p p' 
\ee
\normalsize

\subsubsection{Fourth Order}

The contact potential contribution of order four reads:
\be
V^{(4)}(\vec{p'},\vec{p}) &=&
D_1 q^4 +
D_2 k^4 +
D_3 q^2 k^2 +
D_4 (\vec{q} \times \vec{k})^2 
\nonumber 
\\ &+& 
\left(
D_5 q^4 +
D_6 k^4 +
D_7 q^2 k^2 +
D_8 (\vec{q} \times \vec{k})^2 
\right) \vec{\sigma}_1 \cdot \vec{\sigma}_2 
\nonumber 
\\ &+& 
\left(
D_9 q^2 +
D_{10} k^2 
\right) \left( -i \vec{S} \cdot (\vec{q} \times \vec{k}) \right)
\nonumber 
\\ &+& 
\left(
D_{11} q^2 +
D_{12} k^2 
\right) ( \vec{\sigma}_1 \cdot \vec{q} )\,( \vec{\sigma}_2 
\cdot \vec{q})
\nonumber 
\\ &+& 
\left(
D_{13} q^2 +
D_{14} k^2 
\right) ( \vec{\sigma}_1 \cdot \vec{k} )\,( \vec{\sigma}_2 
\cdot \vec{k})
\nonumber 
\\ &+& 
D_{15} \left( 
\vec{\sigma}_1 \cdot (\vec{q} \times \vec{k}) \, \,
\vec{\sigma}_2 \cdot (\vec{q} \times \vec{k}) 
\right)
\label{eq_ct4}
\ee
The rather lengthy partial-wave expressions of this order
have been relegated to Appendix B.

\subsection{Constructing a Chiral $NN$ Potential
\label{sec_pot}} 

\subsubsection{Conceptual Questions}

The two-nucleon system is non-perturbative as evidenced by the
presence of a shallow bound state (the deuteron)
and large scattering lengths.
Weinberg~\cite{Wei90} showed that the strong enhancement of the
scattering amplitude arises from purely nucleonic intermediate
states. He therefore suggested to use perturbation theory to
calculate the $NN$ potential and to apply this potential
in a scattering equation 
to obtain the $NN$ amplitude. We adopt 
this prescription.

Since the irreducible diagrams that make up the potential
are calculated using
covariant perturbation theory (cf.~Sec.~\ref{sec_pi}), 
it is consistent to
start from the covariant
Bethe-Salpeter (BS) equation~\cite{SB51} describing
two-nucleon scattering.
In operator notation, the BS equation reads
\begin{equation}
T = {\cal V+V\,G} \,T
\label{eq_BS}
\end{equation}
with ${T}$ the invariant amplitude for the two-nucleon scattering process,
${\cal V}$ the sum of all connected two-particle irreducible diagrams, and 
${\cal G}$ the relativistic two-nucleon propagator. 
The BS equation
is equivalent to a set of two equations
\begin{eqnarray}
{T}&=&{V}+{V} \, g \, {T}
\label{eq_bbs}
\\
{V}&=&{\cal V + V\,(G}-g)\,{V} 
\label{eq_BS2}
\\
   &=& {\cal V} + {\cal V}_{1\pi}\,({\cal G}-g)\,{\cal V}_{1\pi} 
    + \ldots \, ,
\label{eq_BS3}
\end{eqnarray}
where $g$ is a covariant three-dimensional propagator
which preserves relativistic elastic unitarity.
We choose the propagator $g$ proposed by 
Blankenbecler and Sugar (BbS)~\cite{BS66}
(for more details on relativistic three-dimensional
reductions of the BS equation, see Ref.~\cite{Mac89}). 
The ellipsis in Eq.~(\ref{eq_BS3}) stands for 
terms of irreducible $3\pi$ and higher pion exchanges
which we neglect.

Note that when we speak of covariance in conjunction with
(heavy baryon) ChPT, we are not referring to manifest
covariance. Relativity and relativistic off-shell
effects are accounted for in terms of
a $Q/M_N$ expansion up to the given order.
Thus, Eq.~(\ref{eq_BS3}) is evaluated in the following way,
\begin{equation}
V  \approx
   {\cal V} (\mbox{\rm on-shell}) +
 {\cal V}_{1\pi}\,{\cal G}\,{\cal V}_{1\pi}-{V}_{1\pi}\,{g}\,{V}_{1\pi} 
\,,
\label{eq_BS4}
\end{equation}
where the pion-exchange content of
   ${\cal V} (\mbox{\rm on-shell})$ is $V_{1\pi} + V_{2\pi}'$
with $V_{1\pi}$ the on-shell 1PE given 
in Eq.~(\ref{eq_1pi}) and $V_{2\pi}'$  the
irreducible $2\pi$ exchanges calculated in Sec.~\ref{sec_pi},
{\it but without the box}. ${\cal V}_{1\pi}$
denotes the relativistic (off-shell) 1PE.
Notice that the term
$({\cal V}_{1\pi}\,{\cal G}\,{\cal V}_{1\pi}-{V}_{1\pi}\,{g}\,{V}_{1\pi})$
represents what has been called ``the (irreducible part of the)
box diagram contribution''
in Sec.~\ref{sec_pi} where it was evaluated at various orders. 

The full chiral $NN$ potential $V$ is given by
irreducible pion exchanges $V_\pi$ 
and contact terms $V_{\rm ct}$,
\begin{equation}
{V} =
V_\pi + 
V_{\rm ct} 
%\, .
\label{eq_pot1}
\end{equation}
with
\begin{equation}
V_\pi = V_{1\pi} + V_{2\pi} + \ldots \,,
\end{equation}
where the ellipsis denotes irreducible $3\pi$ 
and higher pion exchanges which are omitted.
Two-pion exchange contributions appear in various orders
\begin{equation}
V_{2\pi} = 
V_{2\pi}^{(2)} +
V_{2\pi}^{(3)} +
V_{2\pi}^{(4)} 
+ \ldots 
\end{equation}
as calculated in Sec.~\ref{sec_pi}.
Contact terms come in even orders,
\begin{equation}
V_{\rm ct} =
V_{\rm ct}^{(0)} + 
V_{\rm ct}^{(2)} + 
V_{\rm ct}^{(4)} 
+ \ldots
\end{equation}
and were presented in Sec.~\ref{sec_ct}.
The potential $V$ is calculated at a given order.
For example, the potential at NNLO includes
2PE up to 
$V_{2\pi}^{(3)}$ and contacts up to
$V_{\rm ct}^{(2)}$. 
At N$^3$LO, contributions up to
$V_{2\pi}^{(4)}$ and
$V_{\rm ct}^{(4)}$ 
are included.

The potential $V$ satisfies the relativistic
BbS equation, Eq.~(\ref{eq_bbs}).
Defining
\begin{equation}
\widehat{V}({\vec p}~',{\vec p})
\equiv 
\frac{1}{(2\pi)^3}
\sqrt{\frac{M_N}{E_{p'}}}\:  
{V}({\vec p}~',{\vec p})\:
 \sqrt{\frac{M_N}{E_{p}}}
\label{eq_minrel1}
\end{equation}
and
\begin{equation}
\widehat{T}({\vec p}~',{\vec p})
\equiv 
\frac{1}{(2\pi)^3}
\sqrt{\frac{M_N}{E_{p'}}}\:  
{T}({\vec p}~',{\vec p})\:
 \sqrt{\frac{M_N}{E_{p}}}
\label{eq_minrel2}
\end{equation}
with $E_p\equiv \sqrt{M_N^2 + {\vec p}^2}$
(the factor $1/(2\pi)^3$ is added for convenience),
the BbS equation collapses into the usual, nonrelativistic
Lippmann-Schwinger (LS) equation,
\begin{equation}
 \widehat{T}({\vec p}~',{\vec p})= \widehat{V}({\vec p}~',{\vec p})+
\int d^3p''\:
\widehat{V}({\vec p}~',{\vec p}~'')\:
\frac{M}
{{ p}^{2}-{p''}^{2}+i\epsilon}\:
\widehat{T}({\vec p}~'',{\vec p}) \, .
\label{eq_LS}
\end{equation}

Since $\widehat V$ satisfies Eq.~(\ref{eq_LS}), 
it can be used like a usual nonrelativistic potential,
and $\widehat{T}$ is the conventional nonrelativistic 
T-matrix.

Iteration of $\widehat V$ in the LS equation
requires cutting $\widehat V$ off for high momenta to avoid infinities,
This is consistent with the fact that ChPT
is a low-momentum expansion which
is valid only for momenta $Q \ll \Lambda_\chi \approx 1$ GeV.
Thus, we multiply $\widehat V$
with a regulator function
\begin{eqnarray}
{\widehat V}(\vec{ p}~',{\vec p})& 
\longmapsto&
{\widehat V}(\vec{ p}~',{\vec p})
\;\mbox{\boldmath $e$}^{-(p'/\Lambda)^{2n}}
\;\mbox{\boldmath $e$}^{-(p/\Lambda)^{2n}}
\label{eq_regulator}
\\
&&
\approx
{\widehat V}(\vec{ p}~',{\vec p})
\left\{1-\left[\left(\frac{p'}{\Lambda}\right)^{2n}
+\left(\frac{p}{\Lambda}\right)^{2n}\right]+ \ldots \right\} 
\label{eq_reg_exp}
\end{eqnarray}
with the `cutoff parameter' $\Lambda$ around 0.5 GeV.
Equation~(\ref{eq_reg_exp}) provides an indication of the fact that
the exponential cutoff does not necessarily
affect the given order at which 
the calculation is conducted.
For sufficiently large $n$, the regulator introduces contributions that 
are beyond the given order. Assuming a good rate
of convergence of the chiral expansion, such orders are small 
as compared to the given order and, thus, do not
affect the accuracy at the given order.
In our calculations we use, of course,
the full exponential, Eq.~(\ref{eq_regulator}),
and not the expansion. On a similar note, we also
do not expand the square-root factors
in Eqs.~(\ref{eq_minrel1}-\ref{eq_minrel2})
because they are kinematical factors which guarantee
relativistic elastic unitarity.

\begin{table}[t]
\caption{$\chi^2$/datum for the reproduction of the 1999 $np$ 
database~\cite{note2}
by families of $np$ potentials at NLO and NNLO constructed by the
Juelich group~\cite{EGM04}.
\label{tab_chi2a}}
\smallskip
\begin{tabular*}{\textwidth}{@{\extracolsep{\fill}}cccc}
\hline 
\hline 
\noalign{\smallskip}
 $T_{\rm lab}$ bin &  \# of $np$ & 
\multicolumn{2}{c}{\it --- Juelich $np$ potentials --- }\\
 (MeV) 
 & data 
 & NLO
 & NNLO 
\\
\hline 
\hline 
\noalign{\smallskip}
0--100&1058&4--5&1.4--1.9\\ 
100--190&501&77--121&12--32\\ 
190--290&843&140--220&25--69\\ 
\hline 
\noalign{\smallskip}
0--290&2402&67--105&12--27
\\ 
\hline 
\hline 
\end{tabular*}
\end{table}

\begin{table}
\caption{Number of parameters needed for fitting the $np$ data
in phase-shift analysis and by a high-precision $NN$ potential
{\it versus} the number of $NN$ contact terms of EFT based potentials 
at different orders. 
\label{tab_par}}
\smallskip
\begin{tabular*}{\textwidth}{@{\extracolsep{\fill}}ccc|ccc}
\hline 
\hline 
\noalign{\smallskip}
     & Nijmegen     & CD-Bonn        & 
               \multicolumn{3}{c}{\it --- Contact Potentials --- }\\
     & partial-wave & high-precision & $Q^0$ & $Q^2$  & $Q^4$ \\
     & analysis~\cite{Sto93} & potential~\cite{Mac01} & 
                                 LO & NLO/NNLO  & N$^3$LO \\
\hline 
\hline 
\noalign{\smallskip}
$^1S_0$         & 3 & 4 & 1&2 & 4 \\
$^3S_1$         & 3 & 4 & 1&2 & 4 \\
\hline
\noalign{\smallskip}
$^3S_1$-$^3D_1$ & 2 & 2 & 0&1 & 3 \\
\hline
\noalign{\smallskip}
$^1P_1$         & 3 & 3 & 0&1 & 2 \\
$^3P_0$         & 3 & 2 & 0&1 & 2 \\
$^3P_1$         & 2 & 2 & 0&1 & 2 \\
$^3P_2$         & 3 & 3 & 0&1 & 2 \\
\hline
\noalign{\smallskip}
$^3P_2$-$^3F_2$ & 2 & 1 & 0&0 & 1 \\
\hline
\noalign{\smallskip}
$^1D_2$         & 2 & 3 & 0&0 & 1 \\
$^3D_1$         & 2 & 1 & 0&0 & 1 \\
$^3D_2$         & 2 & 2 & 0&0 & 1 \\
$^3D_3$         & 1 & 2 & 0&0 & 1 \\
\hline
\noalign{\smallskip}
$^3D_3$-$^3G_3$ & 1 & 0 & 0&0 & 0 \\
\hline
\noalign{\smallskip}
$^1F_3$         & 1 & 1 & 0&0 & 0 \\
$^3F_2$         & 1 & 2 & 0&0 & 0 \\
$^3F_3$         & 1 & 2 & 0&0 & 0 \\
$^3F_4$         & 2 & 1 & 0&0 & 0 \\
\hline
\noalign{\smallskip}
$^3F_4$-$^3H_4$ & 0 & 0 & 0&0 & 0 \\
\hline
\noalign{\smallskip}
$^1G_4$         & 1 & 0 & 0&0 & 0 \\
$^3G_3$         & 0 & 1 & 0&0 & 0 \\
$^3G_4$         & 0 & 1 & 0&0 & 0 \\
$^3G_5$         & 0 & 1 & 0&0 & 0 \\
\hline
\hline
\noalign{\smallskip}
Total         & 35  & 38 & 2&9 & 24 \\
\hline
\hline
\noalign{\smallskip}
\end{tabular*}
\end{table}

\begin{figure}[t]
\vspace*{-1.2cm}
\hspace*{-1.5cm}
\scalebox{0.40}{\includegraphics{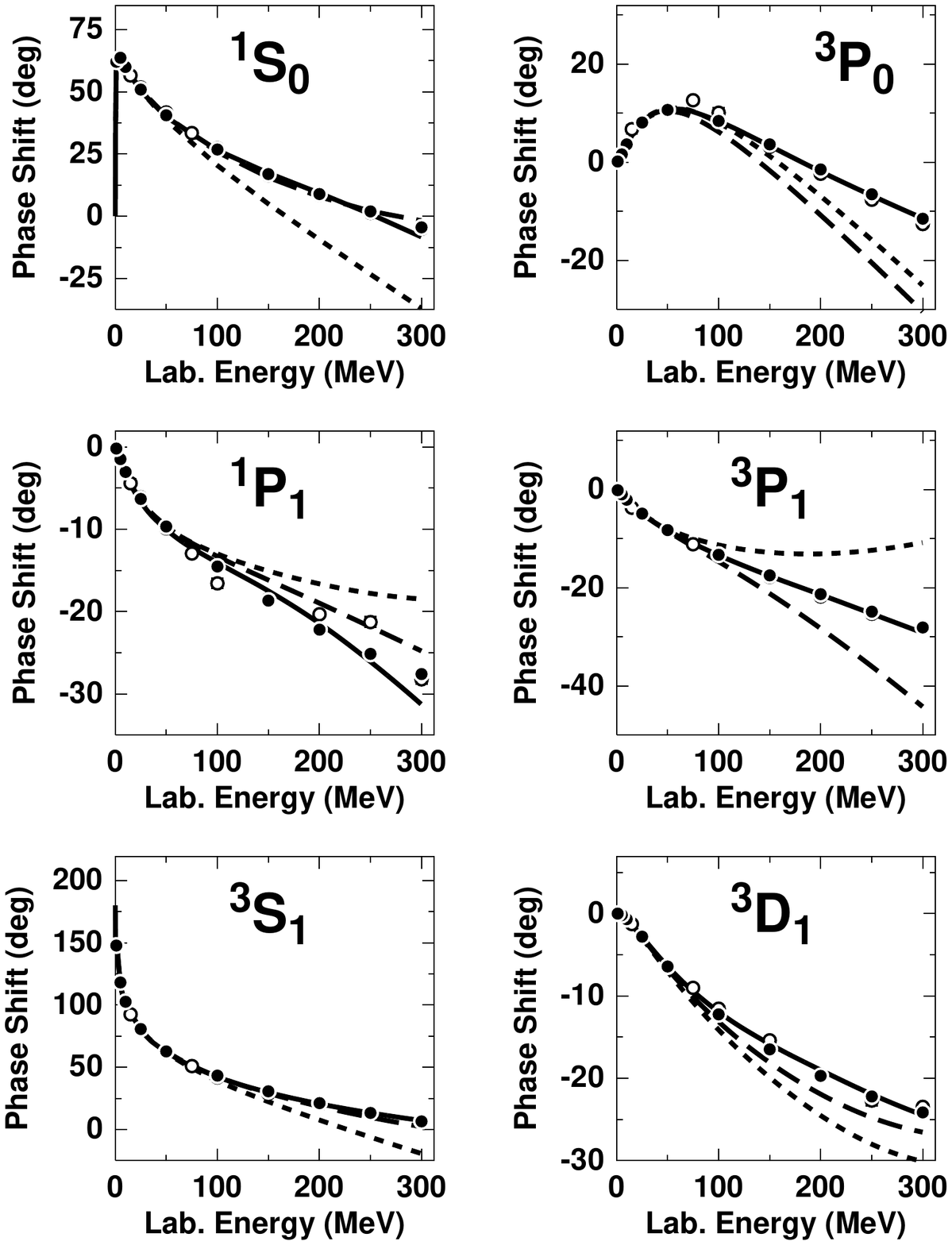}}
\hspace*{-2.3cm}
\scalebox{0.40}{\includegraphics{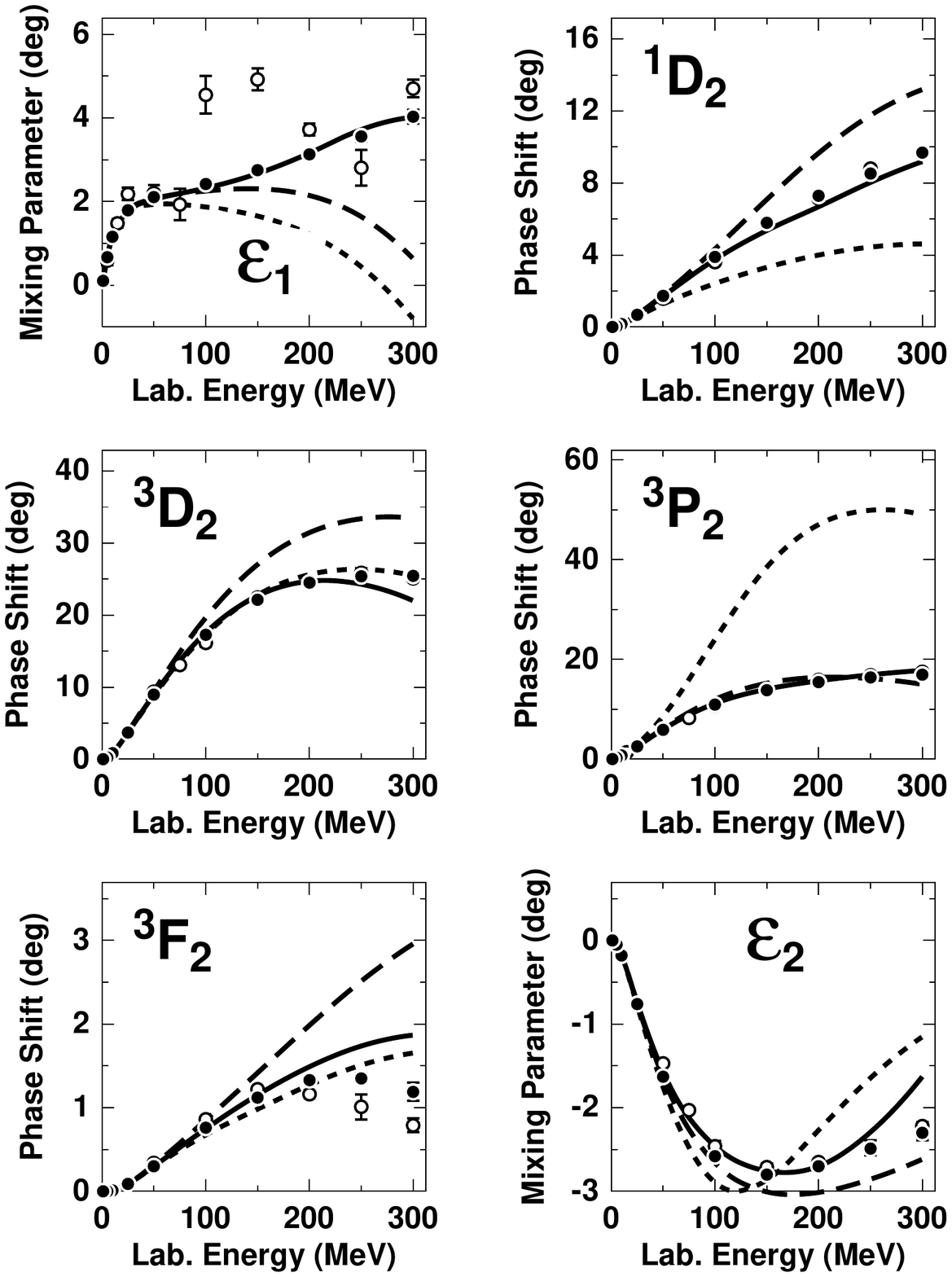}}
\vspace*{-2.3cm}
\caption{\small Phase parameters for $np$ scattering
as calculated from $NN$ potentials at different
orders of ChPT. The dotted line is NLO~\cite{EGM04},
the dashed NNLO~\cite{EGM04},
and the solid N$^3$LO~\cite{EM03}.
Partial waves with total angular momentum $J\leq 2$
are displayed.
Solid dots 
represent the Nijmegen
multienergy $np$ phase shift analysis~\cite{Sto93} and 
open circles are the
GWU/VPI single-energy $np$ analysis SM99~\cite{SM99}.
\label{fig_phorders}}
\end{figure}

\subsubsection{What Order?}

Since in nuclear EFT we are dealing with a perturbative expansion,
at some point, we have to raise the question, to what order 
of ChPT we have to go
to obtain the precision we need.
To discuss this issue on firm grounds, we show in Table~\ref{tab_chi2a}
the $\chi^2$/datum for the fit of the world $np$ data
below 290 MeV for a family of $np$ potentials at 
NLO and NNLO. 
The NLO potentials produce the very large $\chi^2$/datum between 67 and 105,
and the NNLO are between 12 and 27.
The rate of improvement from one order to the other
is very encouraging, but the quality of the reproduction
of the $np$ data at NLO and NNLO is obviously
insufficient for reliable predictions.

Based upon these facts, it has been pointed out in 2002 by
Entem and Machleidt~\cite{EM02a,EM02} that one has
to proceed to N$^3$LO. Consequently, the first N$^3$LO  potential was
published in 2003~\cite{EM03}. 

At N$^3$LO, there are 24 contact terms (24 parameters)
which contribute to the partial waves with $L\leq 2$ 
(cf.~Sec.~\ref{sec_ct}).
In Table~\ref{tab_par}, column `$Q^4$/N$^3$LO', we show how these
terms/parameters are distributed over the various partial waves.
For comparison, we also show the number of parameters
used in the Nijmegen partial wave analysis (PWA93)~\cite{Sto93}
and in the high-precision CD-Bonn potential~\cite{Mac01}.
The table reveals that, for $S$ and $P$ waves, 
the number of parameters
used in high-precision phenomenology and in EFT at N$^3$LO
are about the same.
Thus, the EFT approach provides retroactively a justification
for what the phenomenologists of the 1990's were doing.
At NLO and NNLO, the number of parameters is substantially
smaller than for PWA93 and CD-Bonn, which explains why
these orders are insufficient for a quantitative potential.
This fact is also clearly reflected in Fig.~\ref{fig_phorders}
where phase shifts are shown for potentials constructed
at NLO, NNLO, and N$^3$LO.

\begin{table}[t]
\caption{$\chi^2$/datum for the reproduction of the 1999 
{\boldmath\bf $np$ database}~\cite{note2}
by various $np$ potentials.
(Numbers in parentheses are the values of cutoff parameters 
in units of MeV
used in the regulators of the chiral potentials.)
\label{tab_chi2b}}
\smallskip
\begin{tabular*}{\textwidth}{@{\extracolsep{\fill}}ccccc}
%\begin{tabular}{cc|c|c|c}
\hline 
\hline 
\noalign{\smallskip}
 $T_{\rm lab}$ bin
 & \# of {\boldmath $np$}
 & {\it Idaho}
 & {\it Juelich}
 & Argonne         
\\
 (MeV)
 & data
 & N$^3$LO~\cite{EM03}
 & N$^3$LO~\cite{EGM05} 
 & $V_{18}$~\cite{WSS95}
\\
 & 
 & (500--600)
 & (600/700--450/500)
 & 
\\
\hline 
\hline 
\noalign{\smallskip}
0--100&1058&1.0--1.1&1.0--1.1&0.95\\ 
100--190&501&1.1--1.2&1.3--1.8&1.10\\ 
190--290&843&1.2--1.4&2.8--20.0&1.11\\ 
\hline 
\noalign{\smallskip}
0--290&2402&1.1--1.3&1.7--7.9&1.04
\\ 
\hline 
\hline 
\end{tabular*}
\end{table}

\begin{table}[t]
\caption{$\chi^2$/datum for the reproduction of the 1999 
{\boldmath\bf $pp$ database}~\cite{note2}
by various $pp$ potentials. Notation as in Fig.~\ref{tab_chi2b}.
\label{tab_chi2c}}
\smallskip
\begin{tabular*}{\textwidth}{@{\extracolsep{\fill}}ccccc}
\hline 
\hline 
\noalign{\smallskip}
 $T_{\rm lab}$ bin
 & \# of {\boldmath $np$}
 & {\it Idaho}
 & {\it Juelich}
 & Argonne         
\\
 (MeV)
 & data
 & N$^3$LO~\cite{EM03}
 & N$^3$LO~\cite{EGM05} 
 & $V_{18}$~\cite{WSS95}
\\
 &  
 & (500--600)
 & (600/700--450/500)
 & 
\\
\hline 
\hline 
\noalign{\smallskip}
0--100&795&1.0--1.7&1.0--3.8&1.0 \\ 
100--190&411&1.5--1.9&3.5--11.6&1.3 \\ 
190--290&851&1.9--2.7&4.3--44.4&1.8 \\ 
\hline 
\noalign{\smallskip}
0--290&2057&1.5--2.1&2.9--22.3&1.4 
\\ 
\hline 
\hline 
\end{tabular*}
\end{table}

\subsubsection{Charge-Dependence}

For an accurate fit of the low-energy $pp$ and $np$ data, 
charge-dependence is important.
We include charge-dependence up to next-to-leading order 
of the isospin-violation scheme 
(NL\O, in the notation of Ref.~\cite{WME01}).
Thus, we include
the pion mass difference in 1PE and the Coulomb potential
in $pp$ scattering, which takes care of the L\O\/ contributions. 
At order NL\O, we have the pion mass difference in 2PE at NLO,
$\pi\gamma$ exchange~\cite{Kol98}, and two charge-dependent
contact interactions of order $Q^0$ which make possible
an accurate fit of the 
three different $^1S_0$ scattering 
lengths, $a_{pp}$, $a_{nn}$, and $a_{np}$.

\begin{figure}[t]
\vspace*{-1.2cm}
\hspace*{-1.5cm}
\scalebox{0.40}{\includegraphics{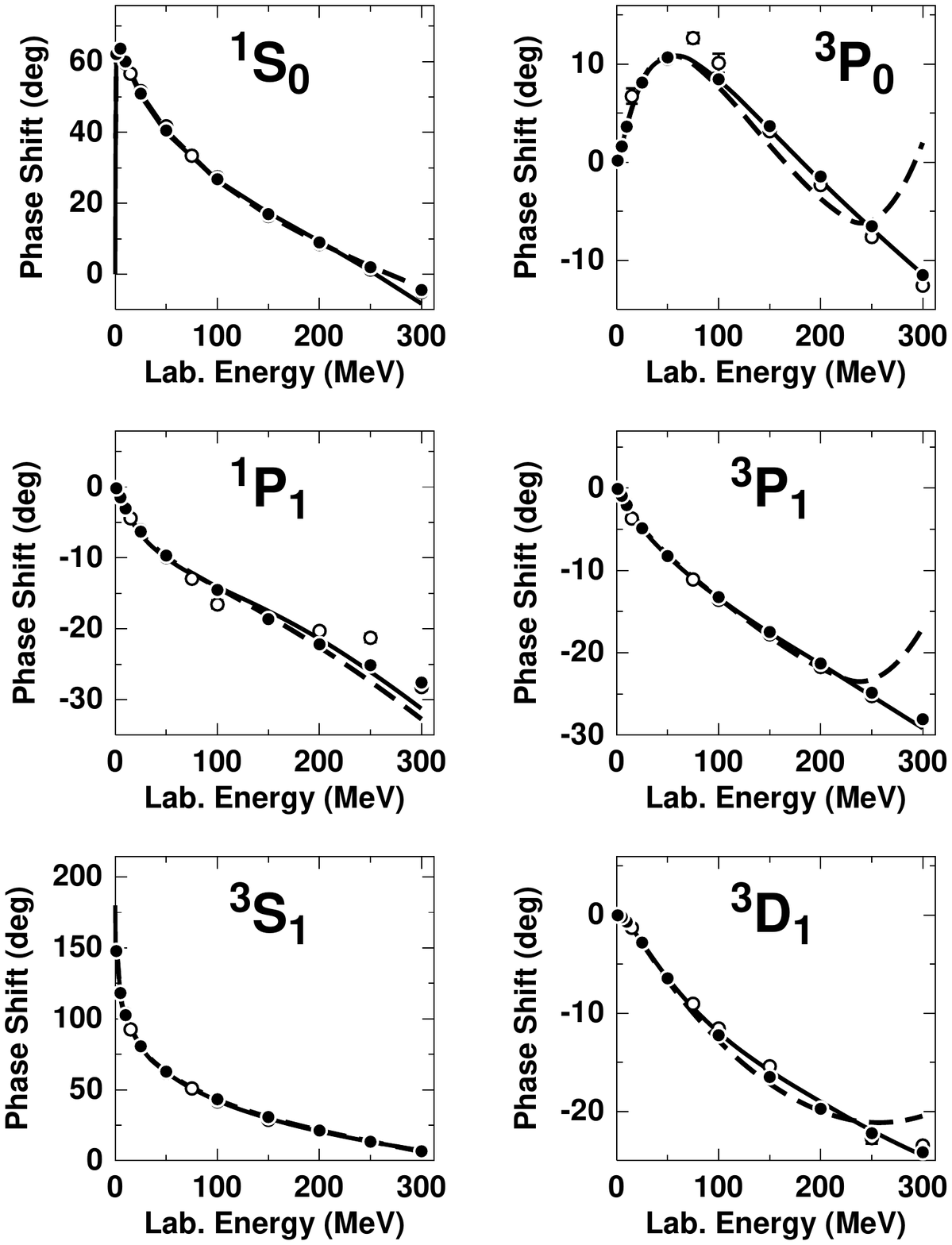}}
\hspace*{-2.3cm}
\scalebox{0.40}{\includegraphics{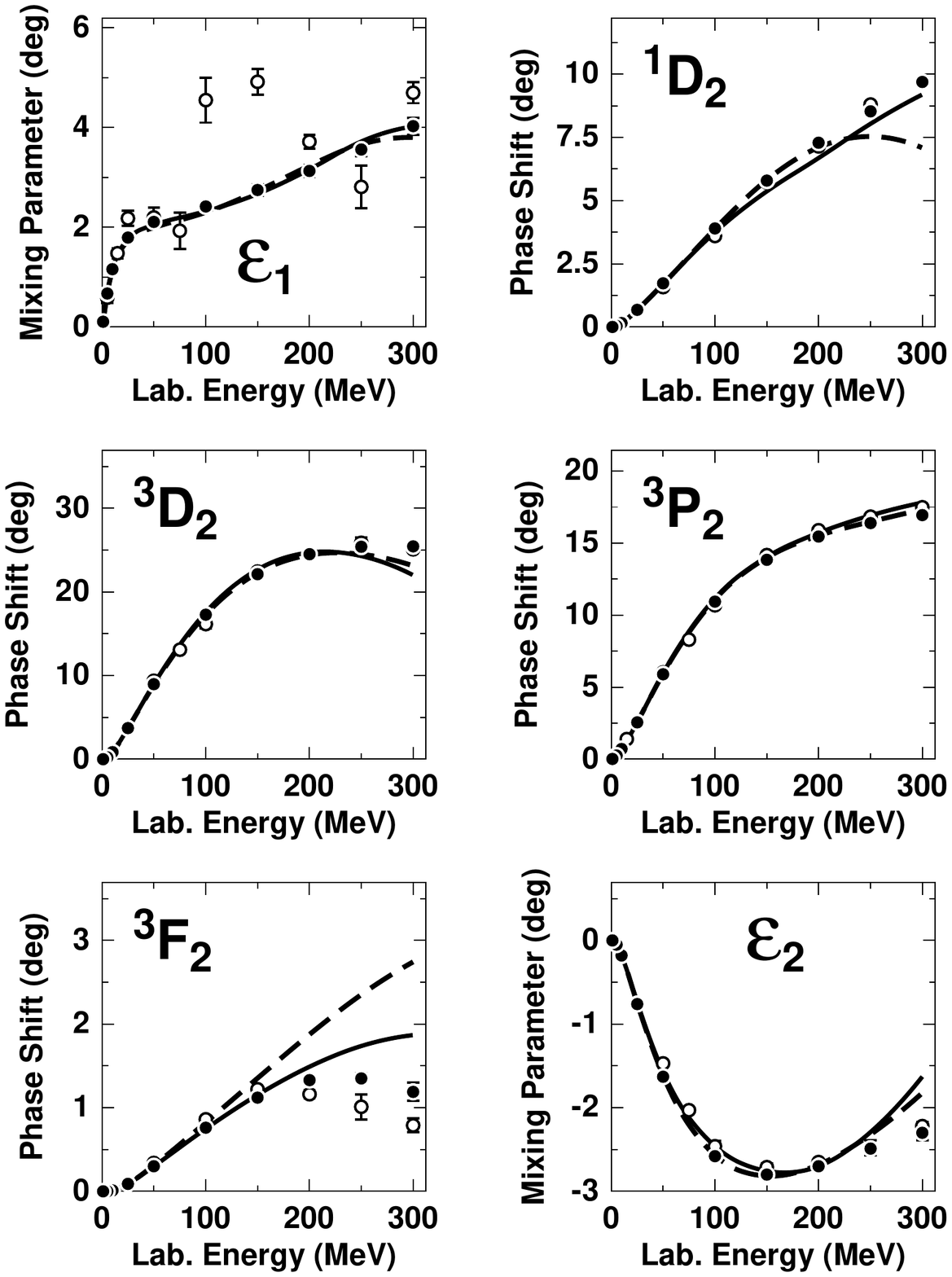}}
\vspace*{-2.3cm}
\caption{\small Neutron-proton phase parameters as described by
two potentials at N$^3$LO.
The solid curve is calculated from
the Idaho N$^3$LO potential~\cite{EM03}
while the dashed curve is from
the Juelich~\cite{EGM05} one.
Solid dots and open circles as in Fig.~\ref{fig_phorders}.
\label{fig_phep}}
\end{figure}

\subsubsection{A Quantitative $NN$ Potential at N$^3$LO
\label{sec_potn3lo}}

\paragraph{$NN$ Scattering.}

The fitting procedure starts with 
the peripheral partial waves because they depend on
fewer parameters.
Partial waves with $L\geq 3$
are exclusively determined by 1PE and 2PE because
the N$^3$LO contacts contribute to $L\leq 2$ only.
1PE and 2PE at N$^3$LO depend on
the axial-vector coupling constant, $g_A$ (we use $g_A=1.29$),
the pion decay constant, $f_\pi=92.4$ MeV,
and eight low-energy constants (LECs) that appear in the 
dimension-two and dimension-three $\pi N$ Lagrangians,
Eqs.~(\ref{eq_L2ct}) and (\ref{eq_L3rel}).
In the fitting process,
we varied three of them, namely, $c_2$, $c_3$, and $c_4$.
We found that the other LECs are not very effective in the
$NN$ system and, therefore, we kept them at the
values determined from $\pi N$ (cf.\ Table~\ref{tab_LEC}).
The most influential constant is $c_3$, which has to be chosen
on the low side (slightly more than one standard deviation
below its $\pi N$ determination) for an optimal fit of the $NN$ data. 
As compared to a calculation that strictly uses the $\pi N$
values for $c_2$ and $c_4$, our choices for these two LECs 
lower the $^3F_2$ and $^1F_3$ phase shifts bringing them 
into closer agreement with the phase shift analysis. The other
$F$ waves and the higher partial waves are essentially unaffected
by our variations of $c_2$ and $c_4$. Overall, the fit
of all $J\geq 3$ waves is very good.

We turn now to the lower partial waves.
Here, the most important fit parameters are the ones associated
with the 24 contact terms that contribute to the partial waves
with $L\leq 2$. In addition, we have two charge-dependent
contacts which are used to fit the
three different $^1S_0$ scattering 
lengths, $a_{pp}$, $a_{nn}$, and $a_{np}$.

In the optimization procedure, we fit first phase shifts,
and then we refine the fit by minimizing the
$\chi^2$ obtained from a direct comparison with the data.
The $\chi^2/$datum for the fit of the $np$ data below
290 MeV is shown in Table~\ref{tab_chi2b}, 
and the corresponding one for $pp$
is given in Table~\ref{tab_chi2c}.
These tables show that at N$^3$LO
a $\chi^2$/datum comparable to the high-precision
Argonne $V_{18}$~\cite{WSS95} potential can, indeed, be achieved.
The ``Idaho'' N$^3$LO potential~\cite{EM03} produces
a $\chi^2$/datum = 1.1 
for the world $np$ data below 290 MeV
which compares well with the $\chi^2$/datum = 1.04
by the Argonne potential.
In 2005, also the Juelich group produced
several N$^3$LO $NN$ potentials~\cite{EGM05}, the best of which
fits the $np$ data with
a $\chi^2$/datum = 1.7 and the worse with 
a $\chi^2$/datum = 7.9 (see 
Table~\ref{tab_chi2b}).
While 7.9 is clearly unacceptable for any meaningful
application, a $\chi^2$/datum of 1.7 is reasonable,
although it does not meet
the precision standard that few-nucleon physicists 
established in the 1990's.

Turning to $pp$,
the $\chi^2$ for $pp$ data are typically
larger than for $np$
because of the higher precision of $pp$ data.
Thus, the Argonne $V_{18}$ produces
a $\chi^2$/datum = 1.4 for the world $pp$ data
below 290 MeV and the best Idaho N$^3$LO $pp$ potential obtains
1.5. The fit by the best Juelich 
N$^3$LO $pp$ potential results in
a $\chi^2$/datum = 2.9 which, again, 
is not quite consistent with the precision standards established
in the 1990's.
The worst Juelich N$^3$LO $pp$ potential produces
a $\chi^2$/datum of 22.3 and 
is incompatible with reliable predictions.

Phase shifts of $np$ scattering from the best Idaho 
(solid line) and Juelich (dashed line)
N$^3$LO $np$ potentials are shown in Figure~\ref{fig_phep}.
The phase shifts confirm what the corresponding
$\chi^2$ have already revealed.

\paragraph{The Deuteron.}

\begin{table}
\small
\caption{\small Deuteron properties as predicted by various $NN$
potentials are compared to empirical information.
(Deuteron binding energy $B_d$, asymptotic $S$ state $A_S$,
asymptotic $D/S$ state $\eta$, deuteron radius $r_d$,
quadrupole moment $Q$, $D$-state probability $P_D$; the calculated
$r_d$ and $Q$ are without meson-exchange current contributions
and relativistic corrections.)
\label{tab_deu}}
\smallskip
\begin{tabular*}{\textwidth}{@{\extracolsep{\fill}}llllll}
\hline 
\hline 
\noalign{\smallskip}
 & Idaho & Juelich \\
 & N$^3$LO~\cite{EM03} & N$^3$LO~\cite{EGM05} & CD-Bonn\cite{Mac01}
 & AV18\cite{WSS95} & Empirical$^a$ \\
 & (500) & (550/600) \\
\hline
\noalign{\smallskip}
$B_d$ (MeV) &
 2.224575& 2.218279&
 2.224575 & 2.224575 & 2.224575(9) \\
$A_S$ (fm$^{-1/2}$) &
 0.8843& 0.8820 &
0.8846 & 0.8850 & 0.8846(9)  \\
$\eta$         & 
 0.0256& 0.0254&
0.0256& 0.0250&0.0256(4) \\
$r_d$ (fm)   &
 1.975& 1.977&
 1.966 &
 1.967 &
 1.97535(85) \\
$Q$ (fm$^2$) &
 0.275& 0.266&
 0.270 & 
 0.270 &
 0.2859(3)  \\
$P_D$ (\%)    & 
 4.51& 3.28&
4.85 & 5.76  \\
\hline
\hline
\end{tabular*}
\footnotesize
$^a$See Table XVIII of Ref.~\cite{Mac01} for references;
the empirical value for $r_d$ is from Ref.~\cite{Hub98}.\\
\end{table}

The reproduction of the deuteron parameters is shown 
in Table~\ref{tab_deu}.
We present results for two N$^3$LO potentials, namely, Idaho~\cite{EM03}
and Juelich~\cite{EGM05}.
Remarkable are the predictions by the chiral potentials
for the deuteron radius which are in good agreement with the latest
empirical value obtained by the isotope-shift method~\cite{Hub98}. 
All $NN$ potentials of the past 
(Table~\ref{tab_deu} includes two representative examples,
namely, CD-Bonn~\cite{Mac01} and AV18~\cite{WSS95})
fail to reproduce this very precise new value for the deuteron radius.

\begin{figure}[t]
\vspace*{-4cm}
\hspace*{0.5cm}
\scalebox{0.50}{\includegraphics{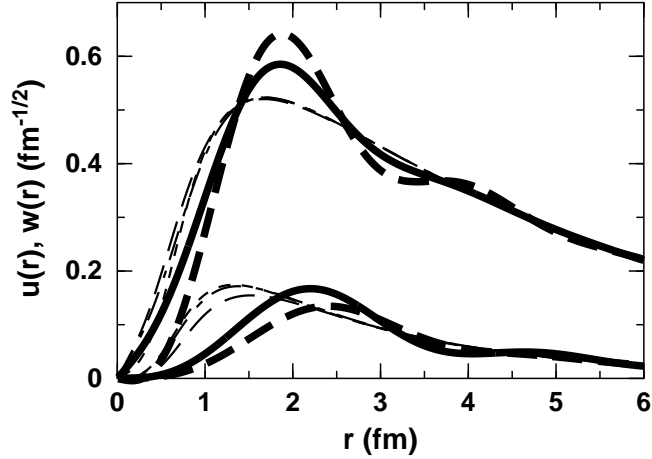}}
\vspace*{-4.0cm}
\caption{Deuteron wave functions: the family of larger curves are $S$-waves, 
the smaller ones $D$-waves. The thick lines represent 
the wave functions derived from chiral $NN$ potentials
at order N$^3$LO (thick solid: Idaho~\cite{EM03}, thick dashed:
Juelich~\cite{EGM05}). The thin dashed, dash-dotted, and dotted lines
refer to the wave functions of the CD-Bonn\protect~\cite{Mac01}, 
Nijm-I\protect~\cite{Sto94}, and AV18\protect~\cite{WSS95} potentials, 
respectively.
\label{fig_deu}}
\end{figure}

In Fig.~\ref{fig_deu}, we display the deuteron wave functions derived from
the N$^3$LO potentials and compare them
with wave functions based upon conventional $NN$ potentials from the
recent past. Characteristic differences are noticeable; in particular,
the chiral wave functions are shifted towards larger $r$ which explains the
larger deuteron radius.

\section{Many-Nucleon Forces}
As noted before, 
an important advantage of the EFT approach to nuclear forces
is that it creates two- and many-nucleon forces on an equal
footing.

\begin{figure}[t]
\vspace*{-5cm}
\hspace*{1.0cm}
\scalebox{0.50}{\includegraphics{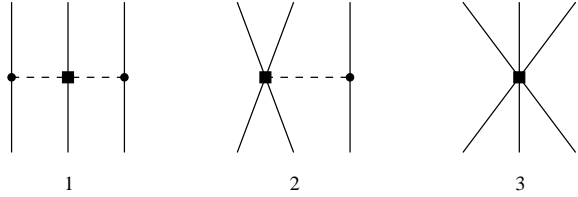}}
\vspace*{-6cm}
\caption{\small The three-nucleon force at NNLO
(from Ref.~\cite{Epe02b}).
\label{fig_3nf_nnlo}}
\end{figure}
 
\subsection{Three-Nucleon Forces}
The first non-vanishing 3NF terms
occur at NNLO and are shown in 
Fig.~\ref{fig_3nf_nnlo} (cf.\ also Fig.~\ref{fig_hi},
row `$Q^3$/NNLO', column `3N Force'). 
There are three diagrams: the 2PE, 1PE,
and 3N-contact interactions~\cite{Kol94,Epe02b}.
The 2PE 3N-potential is given by
\begin{equation}
V^{\rm 3NF}_{\rm 2PE} = 
\left( \frac{g_A}{2f_\pi} \right)^2
\frac12 
\sum_{i \neq j \neq k}
\frac{
( \vec \sigma_i \cdot \vec q_i ) 
( \vec \sigma_j \cdot \vec q_j ) }{
( q^2_i + m^2_\pi )
( q^2_j + m^2_\pi ) } \;
F^{\alpha\beta}_{ijk} \;
\tau^\alpha_i \tau^\beta_j
\label{eq_3nf_nnloa}
\end{equation}
with $\vec q_i \equiv \vec{p_i}' - \vec p_i$, 
where 
$\vec p_i$ and $\vec{p_i}'$ are the initial
and final momenta of nucleon $i$, respectively, and
\begin{equation}
F^{\alpha\beta}_{ijk} = \delta^{\alpha\beta}
\left[ - \frac{4c_1 m^2_\pi}{f^2_\pi}
+ \frac{2c_3}{f^2_\pi} \; \vec q_i \cdot \vec q_j \right]
+ 
\frac{c_4}{f^2_\pi}  
\sum_{\gamma} 
\epsilon^{\alpha\beta\gamma} \;
\tau^\gamma_k \; \vec \sigma_k \cdot [ \vec q_i \times \vec q_j] \; .
\label{eq_3nf_nnlob}
\end{equation}  
The vertex involved in this 3NF term is the two-derivative
$\pi\pi NN$ vertex (solid square in Fig.~\ref{fig_3nf_nnlo}) 
which we encountered
already  in the 2PE contribution to the $NN$ potential at NNLO.
Thus, there are no new parameters and the contribution
is fixed by the LECs used in $NN$.
The 1PE contribution is
\begin{equation}
V^{\rm 3NF}_{\rm 1PE} = 
D \; \frac{g_A}{8f^2_\pi} 
\sum_{i \neq j \neq k}
\frac{\vec \sigma_j \cdot \vec q_j}{
 q^2_j + m^2_\pi }
( \mbox{\boldmath $\tau$}_i \cdot \mbox{\boldmath $\tau$}_j ) 
( \vec \sigma_i \cdot \vec q_j ) 
\label{eq_3nf_nnloc}
\end{equation}
and, finally, the 3N contact term reads
\begin{equation}
V^{\rm 3NF}_{\rm ct} = E \; \frac12
\sum_{j \neq k} 
 \mbox{\boldmath $\tau$}_j \cdot \mbox{\boldmath $\tau$}_k  \; .
\label{eq_3nf_nnlod}
\end{equation}
The last two 3NF terms involve two new vertices
(that do not appear in the 2N problem), namely,
the $\pi NNNN$ vertex with parameter $D$
and a $6N$ vertex with parameters $E$.
To pin them down, one needs two
observables that involve at least three nucleons. 
In Ref.~\cite{Epe02b},
the triton binding energy and the $nd$ doublet scattering
length $^2a_{nd}$ were used.
Alternatively, one may also choose the binding
energies of $^3$H and $^4$He~\cite{Nog06}.
Once $D$ and $E$ are fixed, the results for other
3N, 4N, \ldots  observables are predictions.
In Refs.~\cite{Nog04,Nog06}, the first calculations of the
structure of light nuclei 
($^6$Li and $^7$Li) were reported.
Recently, the structure of nuclei with
$A=10-13$ nucleons has been calculated 
using the {\it ab initio} no-core shell model and
applying chiral two and three-nucleon forces~\cite{Nav07}.
The results are very encouraging.
Concerning the famous `$A_y$ puzzle', the above 3NF terms
yield some improvement of the predicted $nd$ $A_y$, however,
the problem is not solved~\cite{Epe02b}.

\begin{figure}[t]
%\vspace*{-1cm}
\hspace*{1.5cm}
\scalebox{0.45}{\includegraphics{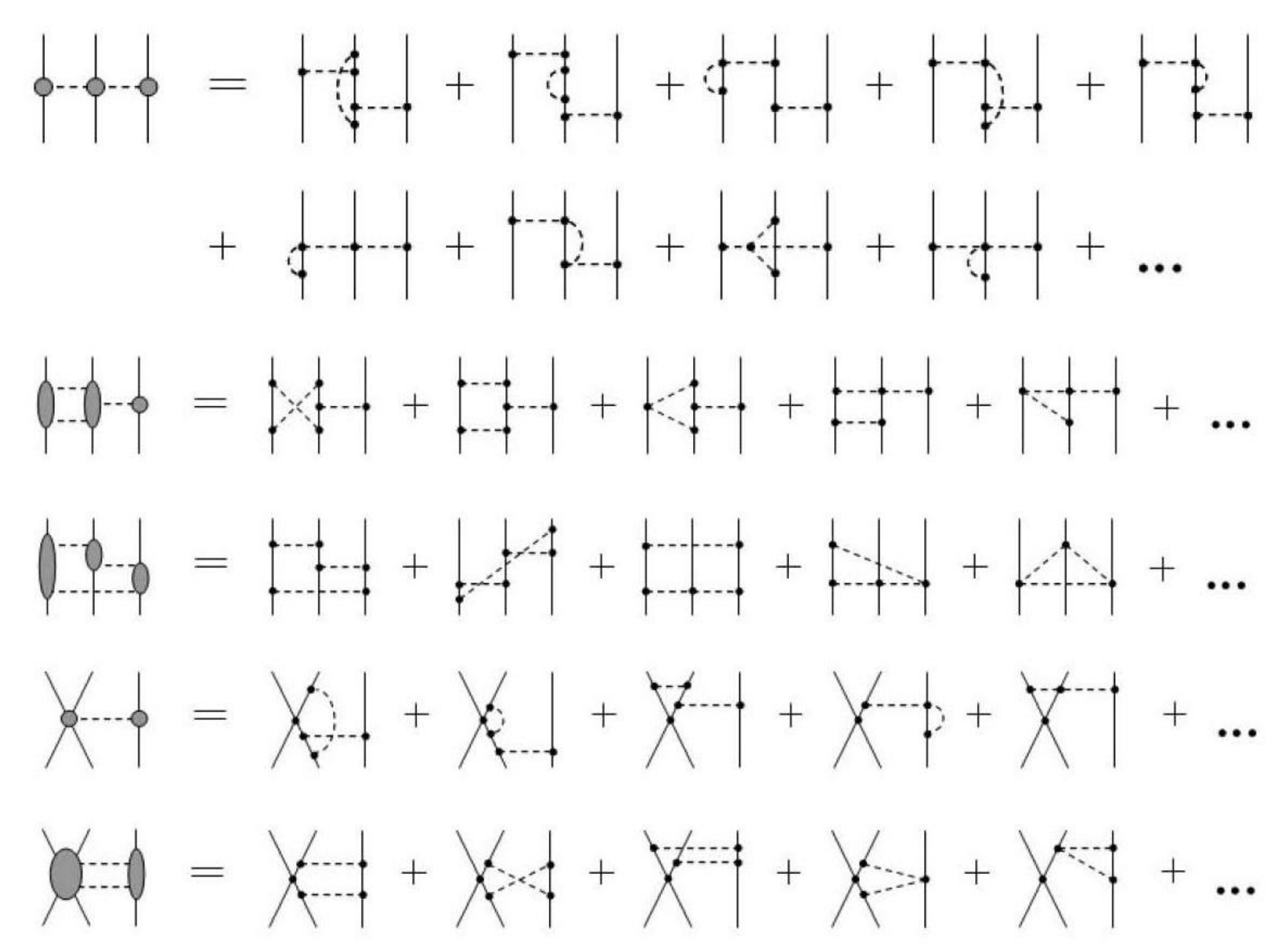}}
\caption{\small Three-nucleon force contributions at 
N$^3$LO (from Ref.~\cite{Mei06}).
\label{fig_3nf_n3lo}}
\end{figure}

Note that the 3NF expressions given in
Eqs.~(\ref{eq_3nf_nnloa})-(\ref{eq_3nf_nnlod}) above are the ones 
that occur at NNLO, and all calculations to date
have included only those.
Since we have to proceed to N$^3$LO for sufficient
accuracy of the 2NF, then consistency requires that
we also consider the 3NF at N$^3$LO.
The 3NF at N$^3$LO is very involved as can be seen
from Fig.~\ref{fig_3nf_n3lo}, but it
does not depend on
any new parameters. It is presently under 
construction~\cite{Mei06}.
So, for the moment, we can only hope that the $A_y$ puzzle
may be solved by a complete calculation at N$^3$LO.

\subsection{Four-Nucleon Forces}

\begin{figure}[t]
\vspace*{-3cm}
\hspace*{2.3cm}
\scalebox{0.35}{\includegraphics{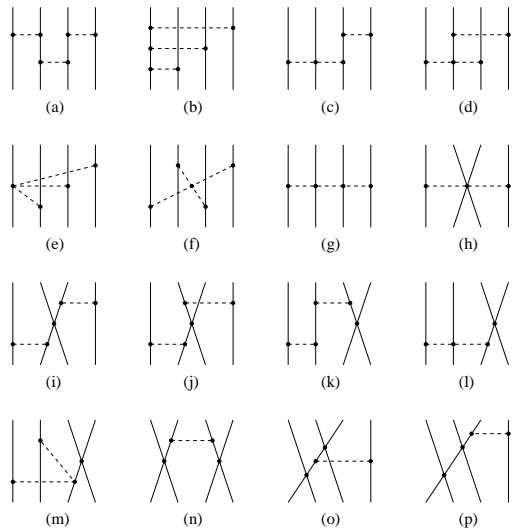}}
\caption{\small The four-nucleon force at N$^3$LO
(from Ref.~\cite{Epe06}).
\label{fig_4nf}}
\end{figure}

In ChPT, four-nucleon forces (4NF) appear for the first
time at N$^3$LO ($\nu = 4$). Thus, N$^3$LO is the leading order
for 4NF. Assuming a good rate of convergence, a contribution
of order $(Q/\Lambda_\chi)^4$ is expected to be rather small.
Thus, ChPT predicts 4NF to be essentially insignificant, 
consistent
with experience. Still, nothing is fully proven in physics
unless we have performed explicit calculations.
Very recently, the first such calculation has been performed:
The chiral 4NF, Fig.~\ref{fig_4nf}, has been applied
in a calculation of the $^4$He binding energy
and found to contribute a few 100 keV~\cite{Roz06}. It should be
noted that this preliminary
calculation involves many approximations,
but it certainly provides the right order of magnitude
of the result, which is
indeed very small
as compared to the full $^4$He binding energy
of 28.3 MeV.

\section{Conclusions}

The theory of nuclear forces has made great progress
since the turn of the millennium.
Nucleon-nucleon potentials have been developed that are based 
on proper theory (EFT for low-energy QCD) 
and are of high-precision, at the same time. 
Moreover, the theory generates
two- and many-body forces on an equal footing
and provides a theoretical explanation for
the empirically known fact that 2NF $\gg$ 3NF $\gg$ 4NF
\ldots.

At N$^3$LO~\cite{EM02,EM03}, the accuracy can be achieved that
is necessary and sufficient for reliable microscopic nuclear 
structure predictions.
First calculations applying the N$^3$LO
$NN$ potential~\cite{EM03} in the conventional shell 
model~\cite{Cor02,Cor05}, the {\it ab initio} no-core shell 
model~\cite{NC04,FNO05,Var05},
the coupled cluster formalism
\cite{Kow04,DH04,Wlo05,Dea05,Gou06},
and the unitary-model-operator approach~\cite{FOS04}
have produced promising results. 

The 3NF at NNLO is known~\cite{Kol94,Epe02b} and has been
applied in few-nucleon reactions~\cite{Epe02b,Erm05,Wit06}
as well as the structure of light nuclei~\cite{Nog04,Nog06,Nav07}. 
However, the famous `$A_y$ puzzle' of nucleon-deuteron
scattering is not resolved by the 3NF at NNLO. Thus,
one important outstanding issue is the 3NF at N$^3$LO,
which is under construction~\cite{Mei06}.

Another open question that needs to be settled 
is whether Weinberg power counting, which is applied in
all current $NN$ potentials, is consistent. This controversial issue 
is presently being debated in the literature~\cite{NTK05,EM06}.
\\
\\

\noindent
{\large\bf Acknowledgements}

It is a pleasure to thank the organizers of this workshop,
particularly, Ananda Santra, for their warm hospitality.
I gratefully acknowledge numerous discussions with
my collaborator D. R. Entem.
This work was supported in part by the U.S. National
Science Foundation under Grant No.~PHY-0099444.

\appendix

\section{Fourth Order Two-Pion Exchange Contributions}

The fourth order 2PE contributions consist of two classes: 
the one-loop (Fig.~\ref{fig_diag2}) and 
the two-loop diagrams (Fig.~\ref{fig_diag3}).

\subsection{One-loop diagrams}

This large pool of diagrams can be analyzed in a systematic
way by introducing the following well-defined
subdivisions.

\subsubsection{$c_i^2$ contributions.}

The only contribution of this kind comes from the football diagram 
with both
vertices proportional to $c_i$ (first row of
Fig.~\ref{fig_diag2}). One obtains~\cite{Kai01a}:
\begin{eqnarray}  
V_C & = & {3 L(q) \over 16 \pi^2 f_\pi^4 } 
\left[
\left( {c_2 \over 6} w^2 +c_3 \widetilde{w}^2 -4c_1 m_\pi^2 \right)^2 
+{c_2^2 \over 45 } w^4 
\right] \,, 
\label{eq_4c2C}
\\
W_T  &=&  -{1\over q^2} W_S 
\nonumber \\
     &=& {c_4^2 w^2 L(q) \over 96 \pi^2 f_\pi^4 } 
\,. 
\label{eq_4c2T}
\end{eqnarray}

\subsubsection{$c_i/M_N$ contributions.}

This class consists of diagrams with one vertex proportional to $c_i$
and one $1/M_N$ correction.
A few graphs that are representative for this class are shown in the
second row of Fig.~\ref{fig_diag2}. 
Symbols with a large solid dot and an open circle denote 
$1/M_N$ corrections of vertices
proportional to $c_i$. They are part of
$\widehat{\cal L}^{(3)}_{\pi N}$, Eq.~(\ref{eq_L3}).
The result for this group of diagrams is~\cite{Kai01a}:
\begin{eqnarray} 
V_C & = & - {g_A^2\, L(q) \over 32 \pi^2 M_N f_\pi^4 } \left[ 
(c_2-6c_3) q^4 +4(6c_1+c_2-3c_3)q^2 m_\pi^2 
\right.  \nonumber \\ && \left.
+6(c_2-2c_3)m_\pi^4
+24(2c_1+c_3)m_\pi^6 w^{-2} \right] \,,
\label{eq_4cMC}
\\
W_C &=& 
-{c_4 q^2 L(q) \over 192 \pi^2 M_N f_\pi^4 } 
\left[ g_A^2 (8m_\pi^2+5q^2) + w^2 \right] 
\,, \\
W_T  &=&  -{1\over q^2} W_S 
\nonumber \\
     &=& -{c_4 L(q) \over 192 \pi^2 M_N f_\pi^4 } 
\left[ g_A^2 (16m_\pi^2+7q^2) - w^2 \right] 
\label{eq_4cMS}
\,,  \\
V_{LS}& = & {c_2 \, g_A^2 \over 8 \pi^2 M_N f_\pi^4 } 
\, w^2 L(q) 
\,, \\
W_{LS}  &=& 
-{c_4 L(q) \over 48 \pi^2 M_N f_\pi^4 } 
\left[ g_A^2 (8m_\pi^2+5q^2) + w^2 \right] 
\,.
\label{eq_4cMLS}
\end{eqnarray}

\subsubsection{$1/M_N^2$ corrections.}

These are relativistic $1/M_N^2$ corrections of the leading order
$2\pi$ exchange diagrams. Typical examples
for this large class are shown in row 3--6 of
Fig.~\ref{fig_diag2}. 
This time, there is no correction from the iterated 1PE, 
Eq.~(\ref{eq_2piitKBW}) or
Eq.~(\ref{eq_2piitEM}),
since the expansion of the factor $M^2_N/E_p$ does not create
a term proportional to $1/M^2_N$. 
The total result for this class is~\cite{Kai01b},
\begin{eqnarray} 
V_C &=& -{g_A^4 \over 32\pi^2 M_N^2 f_\pi^4}
\Bigg[ 
L(q) \, \Big(2m_\pi^8 w^{-4}+8m_\pi^6 w^{-2} -q^4 -2m_\pi^4\Big)
+{ m_\pi^6 \over 2 w^{2}}\, 
\Bigg]
% \,,
\label{eq_4M2C}
\\
W_C           &=& -{1\over 768\pi^2 M_N^2 f_\pi^4} \Bigg\{ L(q) 
\, \bigg[
                8g_A^2 \, \bigg({3\over 2} q^4 +3m_\pi^2 q^2 
+3m_\pi^4
               -6m_\pi^6 w^{-2} 
\nonumber \\   && 
               -k^2(8m_\pi^2 +5q^2) \bigg)
                + 4g_A^4 
                \bigg(k^2\big(20m_\pi^2+7q^2-16m_\pi^4 w^{-2}\big) 
                +16m_\pi^8 w^{-4}
\nonumber \\   && 
                +12 m_\pi^6 w^{-2} 
-4m_\pi^4q^2w^{-2} -5q^4 -6m_\pi^2 q^2-6m_\pi^4 \bigg) 
%\nonumber \\   && 
               -4k^2 w^2 \, \bigg]
\nonumber \\   && 
                \,+\, {16 g_A^4 m_\pi^6 \over w^{2}} \Bigg\} \,,
\\
V_T &=& -{1\over q^2} V_S
    \; = \; {g_A^4 \, L(q) \over 32\pi^2 M_N^2 f_\pi^4} 
        \bigg(k^2+{5\over 8} q^2 +m_\pi^4 w^{-2} \bigg) \,,
\\
W_T &=& -{1\over q^2} W_S 
    \; = \; { L(q) \over 1536\pi^2 M_N^2 f_\pi^4} 
\Bigg[\, 4 g_A^4\, \bigg( 7m_\pi^2+{17\over 4} q^2 +4m_\pi^4 
w^{-2} \bigg) 
\nonumber \\ &&
                 -\, 32 g_A^2\, \bigg( m_\pi^2+{7\over 16}q^2 \bigg) 
                 + w^2 \, \Bigg] \,, 
\\
V_{LS} &=& {g_A^4 \, L(q) \over 4\pi^2 M_N^2 f_\pi^4} 
\bigg( {11 \over 32} q^2 +m_\pi^4 w^{-2}\bigg) \,,
\\
W_{LS} &=&  { L(q) \over 256 \pi^2 M_N^2 f_\pi^4}
\Bigg[\, 16 g_A^2\, \bigg( m_\pi^2+{3\over 8}q^2\bigg)
\nonumber \\ && 
+ \,\frac43\, g_A^4\, \bigg( 4m_\pi^4 w^{-2}-{11\over 4}q^2 
-9m_\pi^2 \bigg) 
                      -w^2 \, \Bigg] \,,
\\
V_{\sigma L} &=& {g_A^4 \, L(q) \over 32\pi^2 M_N^2 f_\pi^4}\;.
\label{eq_4M2sL}
\end{eqnarray} 

\subsection{Two-loop contributions.}

The two-loop contributions are quite intricate.
In Fig.~\ref{fig_diag3}, we attempt a graphical representation of
this class. The gray disk stands for all one-loop $\pi N$ graphs
which are shown in some detail in the lower part of the figure.
Not all of the numerous graphs are displayed. 
Some of the missing ones
are obtained by permutation of the vertices along the nucleon line,
others by inverting initial and final states.
Vertices denoted by a small dot are from the
leading order $\pi N$ Lagrangian
$\widehat{\cal L}^{(1)}_{\pi N}$,
Eq.~(\protect\ref{eq_L1}), except
for the four-pion vertices which are from 
${\cal L}^{(2)}_{\pi\pi}$, Eq.~(\ref{eq_Lpipi}).
The solid square represents vertices proportional to the LECs
$d_i$ which are introduced by the third order Lagrangian 
${\cal L}^{(3)}_{\pi N}$,
Eq.~(\ref{eq_L3rel}).
The $d_i$ vertices occur actually in one-loop $NN$ diagrams, but
we list them among the two-loop $NN$ contributions because they
are needed to absorb divergences generated by 
one-loop $\pi N$ graphs.
Using techniques from dispersion theory,
Kaiser~\cite{Kai01a} calculated the imaginary parts of the
$NN$ amplitudes, Im $V_\alpha(i\mu)$ and Im $W_\alpha(i\mu)$,
which result from analytic continuation to time-like
momentum transfer $q=i\mu-0^+$ with $\mu\geq 2m_\pi$.
From this,
the momentum-space amplitudes $V_\alpha(q)$ and $W_\alpha(q)$
are obtained
via the subtracted dispersion relations:
\begin{eqnarray} 
V_{C,S}(q) &=& 
-{2 q^6 \over \pi} \int_{2m_\pi}^\infty d\mu \,
{{\rm Im\,}V_{C,S}(i \mu) \over \mu^5 (\mu^2+q^2) }\,, 
\\
V_T(q) &=& 
{2 q^4 \over \pi} \int_{2m_\pi}^\infty d\mu \,
{{\rm Im\,}V_T(i \mu) \over \mu^3 (\mu^2+q^2) }\,, 
\end{eqnarray}
and similarly for $W_{C,S,T}$.

In most cases, the dispersion integrals can be solved
analytically and the following expressions are 
obtained~\cite{EM02}:
\begin{eqnarray}
V_C (q) & = &
 \frac{3g_A^4 \widetilde{w}^2 A(q)}{1024 \pi^2 f_\pi^6}
\left[ ( m_\pi^2 + 2q^2 )
\left( 2m_\pi + \widetilde{w}^2 A(q) \right)
+ 4g_A^2 m_\pi \widetilde{w}^2 \right] ;
\nonumber \\
\label{eq_42lC}
\\
W_C(q) & = & W_C^{(a)}(q) + W_C^{(b)}(q) \,,
\\
\mbox{\rm with \hspace*{1.5cm}} 
\nonumber \\
W_C^{(a)}(q) & = &
\frac{L(q)}{18432 \pi^4 f_\pi^6}
\Bigg\{
192 \pi^2 f_\pi^2
w^2 \bar{d}_3
\left[2g_A^2\widetilde{w}^2-\frac35(g_A^2-1)w^2\right]
\nonumber \\
&&
+\left[6g_A^2\widetilde{w}^2-(g_A^2-1)w^2\right]
\Bigg[
 384\pi^2f_\pi^2
\left(\widetilde{w}^2(\bar{d}_1+\bar{d}_2)+4m_\pi^2\bar{d}_5\right)
\nonumber \\
&&
+L(q)
\left(4m_\pi^2(1+2g_A^2)+q^2(1+5g_A^2)\right)
\nonumber \\ &&
-\left(\frac{q^2}{3}(5+13g_A^2)+8m_\pi^2(1+2g_A^2)\right)
\Bigg]
\Bigg\}
\label{eq_WC}
\\
\mbox{\rm and \hspace*{1.6cm}}
\nonumber \\
W_C^{(b)}(q) & = &
-{2 q^6 \over \pi} \int_{2m_\pi}^\infty d\mu \,
{{\rm Im\,}W_C^{(b)}(i \mu) \over \mu^5 (\mu^2+q^2) }\,, 
\\
\mbox{\rm where\hspace*{1.4cm}}
\nonumber \\
{\rm Im}\, W_C^{(b)}(i\mu)&=& 
-{2\kappa \over 3\mu (8\pi f_\pi^2)^3} 
\int_0^1 dx\, 
\Big[ g_A^2(2m_\pi^2-\mu^2) +2(g_A^2-1)\kappa^2x^2 \Big]
\nonumber \\
&& \times \left\{ 
-\,3\kappa^2x^2 
+6 \kappa x \sqrt{m_\pi^2 +\kappa^2 x^2} \ln{ \kappa x 
+\sqrt{m_\pi^2 
+\kappa^2 x^2}\over  m_\pi}
\right.
\nonumber \\
&&
\left.
+g_A^4\left(\mu^2 -2\kappa^2 x^2 -2m_\pi^2\right) 
%\right.  \nonumber \\ && \left.  \times 
\left[ {5\over 6} +{m_\pi^2\over \kappa^2 x^2} 
-\left( 1 +{m_\pi^2\over \kappa^2 x^2} \right)^{3/2} 
\right. \right. \nonumber \\ &&  \left. \left. \times
\ln{ \kappa x +\sqrt{m_\pi^2 +\kappa^2 x^2}\over  m_\pi} \right] 
\right\};   
\\
V_T(q)  &=& V_T^{(a)}(q) + V_T^{(b)}(q) 
\nonumber \\
        &=& - {1 \over q^2}V_S(q) 
 \; = \; - {1\over q^2}\left(V_S^{(a)}(q)+V_S^{(b)}(q)\right) ,
\\
\mbox{\rm with\hspace*{1.5cm}}
\nonumber \\
V_T^{(a)}(q)  &=& - {1\over q^2}V_S^{(a)}(q) 
              \; = \; -\frac{g_A^2 w^2 L(q)}{32 \pi^2 f_\pi^4}
(\bar{d}_{14} - \bar{d}_{15}) 
\label{eq_VT}
\\
\mbox{\rm and \hspace*{1.5cm}}
\nonumber \\
V_T^{(b)}(q)  &=& - {1\over q^2}V_S^{(b)}(q) 
              \; = \; {2 q^4 \over \pi} \int_{2m_\pi}^\infty d\mu \,
{{\rm Im\,}V_T^{(b)}(i \mu) \over \mu^3 (\mu^2+q^2) } \,, 
\\
\mbox{\rm where\hspace*{1.4cm}}
\nonumber \\ 
{\rm Im}\, V_T^{(b)}(i\mu) &=& 
-{2g_A^6 \kappa^3 \over \mu (8\pi f_\pi^2)^3} 
\int_0^1 dx(1-x^2)
\left[
-{1\over 6}+{m_\pi^2 \over \kappa^2x^2}
\right.  \nonumber \\ && \left.  
%\hspace*{3.0cm}
-\left( 1+{m_\pi^2 \over \kappa^2x^2} \right)^{3/2} 
\ln{ \kappa x +\sqrt{m_\pi^2 +\kappa^2 x^2}\over  m_\pi}
\right] ; \\ 
W_T(q) &=& - {1 \over q^2}W_S(q) 
%\nonumber \\
       = \frac{g_A^4 w^2 A(q)}{2048 \pi^2 f_\pi^6}
\left[ w^2 A(q) + 2m_\pi (1 + 2 g_A^2) \right],
\label{eq_42lT}
\end{eqnarray}
where $\kappa \equiv \sqrt{\mu^2/4-m_\pi^2}$.

Note that the analytic solutions hold modulo polynomials.
We have checked the importance of those contributions where 
we could not find an analytic solution and where, therefore, the
integrations have to be performed numerically. It turns out
that the combined effect on $NN$ phase shifts from
$W_C^{(b)}$, $V_T^{(b)}$, and $V_S^{(b)}$
is smaller than 0.1 deg in $F$ and $G$ waves
and smaller than 0.01 deg in $H$ waves,
at $T_{\rm lab} = 300$ MeV (and less at lower energies). 
This renders these contributions negligible. 
Therefore, we omit
$W_C^{(b)}$, $V_T^{(b)}$, and $V_S^{(b)}$
in the construction of chiral $NN$ potentials
at order N$^3$LO.

In Eqs.~(\ref{eq_WC}) and (\ref{eq_VT}), we use the scale-independent
LECs, $\bar{d}_i$, which are obtained by combining the 
scale-dependent ones, $d_i^r (\lambda)$, 
with the chiral logarithm,
$\ln (m_\pi/\lambda)$, or equivalently $\bar{d}_i = d^r_i(m_\pi)$.
The scale-dependent LECs, $d_i^r (\lambda)$, 
are a consequence of renormalization.
For more details about this issue, see Ref.~\cite{FMS98}.

\section{Partial Wave Decomposition of the Fourth Order
Contact Potential}

The contact potential contribution of order four,
Eq.~(\ref{eq_ct4}), decomposes into partial-waves as
follows.

\be
V^{(4)}(^1 S_0)          &=&  \widehat{D}_{^1 S_0}          
({p'}^4 + p^4) +
                              D_{^1 S_0}          {p'}^2 p^2 
\nonumber 
\\
V^{(4)}(^3 P_0)          &=&        D_{^3 P_0}          
({p'}^3 p + p' p^3) 
\nonumber 
\\
V^{(4)}(^1 P_1)          &=&        D_{^1 P_1}          
({p'}^3 p + p' p^3) 
\nonumber 
\\
V^{(4)}(^3 P_1)          &=&        D_{^3 P_1}          
({p'}^3 p + p' p^3) 
\nonumber 
\\
V^{(4)}(^3 S_1)          &=&  \widehat{D}_{^3 S_1}          
({p'}^4 + p^4) +
                              D_{^3 S_1}          {p'}^2 p^2 
\nonumber 
\\
V^{(4)}(^3 D_1)          &=&        D_{^3 D_1}          
{p'}^2 p^2 
\nonumber 
\\
V^{(4)}(^3 S_1 - ^3 D_1) &=&  \widehat{D}_{^3 S_1 - ^3 D_1} 
p^4             +
                              D_{^3 S_1 - ^3 D_1} {p'}^2 p^2
\nonumber 
\\
V^{(4)}(^1 D_2)          &=&        D_{^1 D_2}          
{p'}^2 p^2 
\nonumber 
\\
V^{(4)}(^3 D_2)          &=&        D_{^3 D_2}          
{p'}^2 p^2 
\nonumber 
\\
V^{(4)}(^3 P_2)          &=&        D_{^3 P_2}          
({p'}^3 p + p' p^3) 
\nonumber 
\\
V^{(4)}(^3 P_2 - ^3 F_2) &=&        D_{^3 P_2 - ^3 F_2} {p'}p^3
\nonumber 
\\
V^{(4)}(^3 D_3)          &=&        D_{^3 D_3}          
{p'}^2 p^2 
\ee

The coefficients in the above expressions are given by:
\footnotesize
\be
\widehat{D}_{^1 S_0} &=& 
             D_1    + \frac{1}{16} D_2    + \frac{1}{4}  D_3 - 
           3 D_5    - \frac{3}{16} D_6    - \frac{3}{4}  D_7 - 
             D_{11} - \frac{1}{4}  D_{12}  
 - \frac{1}{4} D_{13} 
\nonumber \\ &&
- \frac{1}{16} D_{14} 
\nonumber \\
D_{^1 S_0} &=& 
\frac{10}{3} D_1    + \frac{5}{24} D_2    +  \frac{1}{6} D_3 + 
 \frac{2}{3} D_4    -           10 D_5    -  \frac{5}{8} D_6 - 
 \frac{1}{2} D_7    -            2 D_8    - \frac{10}{3} D_{11}  
\nonumber \\ &&
 - \frac{1}{6} D_{12} 
                    -  \frac{1}{6} D_{13} - \frac{5}{24} D_{14} - 
 \frac{2}{3} D_{15}  
\nonumber \\
D_{^3 P_0} &=& 
-\frac{4}{3} D_1    + \frac{1}{12} D_2    -  \frac{4}{3} D_5     + 
\frac{1}{12} D_6    -  \frac{2}{3} D_9    -  \frac{1}{6} D_{10}  + 
 \frac{8}{3} D_{11} +  \frac{1}{3} D_{12} -  \frac{1}{3} D_{13}  
\nonumber \\ &&
 -  \frac{1}{6} D_{14}
\nonumber \\
D_{^1 P_1} &=& 
-\frac{4}{3} D_1    + \frac{1}{12} D_2    +            4 D_5     - 
 \frac{1}{4} D_6    +  \frac{4}{3} D_{11} - \frac{1}{12} D_{14}
\nonumber \\
D_{^3 P_1} &=& 
-\frac{4}{3} D_1    + \frac{1}{12} D_2    -  \frac{4}{3} D_5     + 
\frac{1}{12} D_6    -  \frac{1}{3} D_9    - \frac{1}{12} D_{10}  - 
           2 D_{11} -  \frac{1}{6} D_{12} +  \frac{1}{6} D_{13}   
\nonumber \\ &&
 + \frac{1}{8} D_{14}
\nonumber \\
\widehat{D}_{^3 S_1} &=& 
             D_1    + \frac{1}{16} D_2    +  \frac{1}{4} D_3     + 
	     D_5    + \frac{1}{16} D_6    +  \frac{1}{4} D_7     +  
 \frac{1}{3} D_{11} + \frac{1}{12} D_{12} + \frac{1}{12} D_{13}  
\nonumber \\ &&
 + \frac{1}{48} D_{14}
\nonumber \\
D_{^3 S_1} &=& 
\frac{10}{3} D_1    + \frac{5}{24} D_2    +  \frac{1}{6} D_3     + 
 \frac{2}{3} D_4    + \frac{10}{3} D_5    + \frac{5}{24} D_6     + 
 \frac{1}{6} D_7    +  \frac{2}{3} D_8    + \frac{10}{9} D_{11}  
\nonumber \\ &&
+ \frac{1}{18} D_{12} 
                    + \frac{1}{18} D_{13} + \frac{5}{72} D_{14}  + 
 \frac{2}{9} D_{15}
\nonumber \\
D_{^3 D_1} &=& 
\frac{8}{15} D_1    + \frac{1}{30} D_2    - \frac{2}{15} D_3- 
\frac{2}{15} D_4    + \frac{8}{15} D_5    + \frac{1}{30} D_6- 
\frac{2}{15} D_7    - \frac{2}{15} D_8    
\nonumber \\ &&
+  \frac{2}{5} D_9 - \frac{1}{10} D_{10} 
                    -  \frac{4}{9} D_{11} +  \frac{1}{9} D_{12}+ 
 \frac{1}{9} D_{13} - \frac{1}{36} D_{14} -\frac{16}{45} D_{15}
\nonumber \\
\widehat{D}_{^3 S_1 - ^3 D_1} &=& 
-\frac{2\,{\sqrt{2}}}{3} D_{11}  -  \frac{\sqrt{2}}{6} D_{12}- 
      \frac{\sqrt{2}}{6} D_{13}  - \frac{\sqrt{2}}{24} D_{14}
\nonumber \\
D_{^3 S_1 - ^3 D_1} &=& 
-\frac{14\,{\sqrt{2}}}{9} D_{11}  +      \frac{\sqrt{2}}{18} D_{12}+ 
      \frac{\sqrt{2}}{18} D_{13}  - \frac{7\,{\sqrt{2}}}{72} D_{14}+ 
  \frac{2\,{\sqrt{2}}}{9} D_{15}
\nonumber \\
D_{^1 D_2} &=& 
 \frac{8}{15} D_1    +  \frac{1}{30} D_2    -  \frac{2}{15} D_3    - 
 \frac{2}{15} D_4    -   \frac{8}{5} D_5    -  \frac{1}{10} D_6    + 
  \frac{2}{5} D_7    +   \frac{2}{5} D_8    -  \frac{8}{15} D_{11}  
\nonumber \\ &&
 + \frac{2}{15} D_{12} 
                     +  \frac{2}{15} D_{13} -  \frac{1}{30} D_{14} + 
 \frac{2}{15} D_{15}
\nonumber \\
D_{^3 D_2} &=& 
 \frac{8}{15} D_1    +  \frac{1}{30} D_2    -  \frac{2}{15} D_3- 
 \frac{2}{15} D_4    +  \frac{8}{15} D_5    +  \frac{1}{30} D_6- 
 \frac{2}{15} D_7    -  \frac{2}{15} D_8    
\nonumber \\ &&
+  \frac{2}{15} D_9 - \frac{1}{30} D_{10} 
                     +   \frac{4}{5} D_{11} -   \frac{1}{5} D_{12}- 
  \frac{1}{5} D_{13} +  \frac{1}{20} D_{14} +  \frac{4}{15} D_{15}
\nonumber \\
D_{^3 P_2} &=& 
 -\frac{4}{3} D_1    +  \frac{1}{12} D_2    -   \frac{4}{3} D_5+ 
 \frac{1}{12} D_6    +   \frac{1}{3} D_9    +  \frac{1}{12} D_{10}- 
 \frac{2}{15} D_{11} +  \frac{1}{30} D_{12} 
\nonumber \\ &&
 -  \frac{1}{30} D_{13} + \frac{1}{120} D_{14}
\nonumber \\
D_{^3 P_2 - ^3 F_2} &=& 
   \frac{4\,{\sqrt{6}}}{15} D_{11} - 
      \frac{{\sqrt{6}}}{15} D_{12} + 
      \frac{{\sqrt{6}}}{15} D_{13} - 
        \frac{\sqrt{6}}{60} D_{14}
\nonumber \\
D_{^3 D_3} &=& 
 \frac{8}{15} D_1    +  \frac{1}{30} D_2    -  \frac{2}{15} D_3- 
 \frac{2}{15} D_4    +  \frac{8}{15} D_5    +  \frac{1}{30} D_6- 
 \frac{2}{15} D_7    -  \frac{2}{15} D_8    
\nonumber \\ &&
 -  \frac{4}{15} D_9 + 
 \frac{1}{15} D_{10} -  \frac{2}{15} D_{15}
\ee

\end{document}